\documentclass[11pt,a4paper]{article}
\usepackage[T1]{fontenc}
\usepackage[utf8]{inputenc}
\usepackage{lmodern}
\usepackage[margin=25mm,headheight=14pt]{geometry}
\usepackage{amsmath,amssymb,amsthm,mathtools}
\usepackage{graphicx,xcolor}
\usepackage{tikz}
\usepackage{booktabs}
\usepackage{microtype}
\usepackage{fancyhdr}
\usepackage{needspace}
\usepackage[colorlinks=true,allcolors=blue]{hyperref}
\hypersetup{pdftitle={Charged Schrodinger blocks in 1+1 dimensions from the quartic Casimir and endpoint equations},pdfauthor={Hirohiko Shimada}}

\newcommand{\normalA}{\xi}
\newcommand{\normalB}{\eta}
\newcommand{\framex}{{\widehat{x}_1}}
\newcommand{\framey}{{\widehat{x}_2}}
\newcommand{\intx}{{\widehat{x}}}
\newcommand{\inty}{{\widehat{y}}}

\newcommand{\cF}{\mathcal{F}}
\newcommand{\cP}{\mathcal{P}}
\newcommand{\cN}{\mathcal{N}}
\newcommand{\cI}{\mathcal{I}}
\newcommand{\cK}{\mathcal{K}}
\newcommand{\sH}{\mathsf{H}}
\newcommand{\sD}{\mathsf{D}}
\newcommand{\sK}{\mathsf{K}}
\newcommand{\hyp}[2]{{}_{#1}F_{#2}}
\numberwithin{equation}{section}
\newtheorem{proposition}{Proposition}[section]
\newtheorem{lemma}[proposition]{Lemma}
\title{Charged Schr\"odinger blocks in $1+1$ dimensions\\from the quartic Casimir and endpoint equations}
\author{Hirohiko Shimada\\
\small National Institute of Technology, Tsuyama College\\
\small Tsuyama, Okayama 708-8509, Japan\\
\small Email: \href{mailto:hshimada@tsuyama-ct.ac.jp}{\texttt{hshimada@tsuyama-ct.ac.jp}}}
\date{}

\begin{document}
\maketitle
\begin{abstract}
We construct charge-two Schr\"odinger conformal blocks in one spatial
dimension, independently of any dynamical realization. We consider ordered
vacuum four-point functions of fundamental fields of dimension $1/2$,
with homogeneous Schr\"odinger equations at the earliest and latest
insertions and interactions allowed in the intermediate two-particle
interval. For reduced blocks analytic in signed spatial coordinates and
even under simultaneous spatial inversion, two functions of the time
cross-ratio determine the full spatial dependence: the collinear value
and a rescaled mixed spatial derivative. An exact Kummer lift, built from confluent hypergeometric
functions, gives this reconstruction. An operator identity reduces the quartic Casimir equation,
modulo the endpoint equations, to an equivalent coupled system for this
pair. At generic exchanged dimension, OPE conditions select a unique block
up to normalization for each primary parity. We obtain collinear
${}_4F_3$ functions and sums of Kummer products, including the shadow
branches. The blocks depend on the relative sign of the signed coordinates,
distinguishing piecewise linear worldlines with an inflection from those with
monotone segment velocities. In the Calogero model, we recover the previously derived
bosonic four-point function and obtain its fermionic counterpart, with
an independent integral check. At the diffusion-limited fixed point of
$A+B\to\emptyset$, two blocks of opposite primary parity give the same
collinear unlike-species response. 
Nonnegativity for all positions and ordered times selects the even family,
thereby establishing that the response summed over species contains the
even shadow pair with dimensions $\Delta=1,2$.
The nonnegativity argument demonstrates a concrete constraint
for a Schr\"odinger bootstrap beyond collinear data.
\end{abstract}

\section{Introduction}\label{sec:introduction}
Nonequilibrium dynamics exhibits universality. In domain growth after a quench into an ordered phase and in the decay of particle densities in diffusion-limited reactions, the long-time behavior is characterized by exponents and scaling functions that are independent of microscopic details \cite{Bray,TauberHowardVollmayrLee}. 
At equilibrium critical points, the conformal bootstrap constrains universal properties by combining crossing symmetry of four-point functions with positivity from unitarity \cite{FerraraGrilloGatto,Polyakov1974,PolandRychkovVichi}. 
Stochastic dynamics is not unitary, but many of its response functions to particle injection are probabilities or densities of detection events and are therefore nonnegative \cite{Doi,Peliti}. 
In some nonequilibrium systems, the scaling forms of response functions can be constrained by Schr\"odinger symmetry, an extension of dynamical scaling with exponent $z=2$ \cite{Henkel}. 
Four-point responses describe the response at two later spacetime points to perturbations at two earlier points. 
In phase-ordering kinetics, Henkel and Stoimenov have derived scaling relations and asymptotic laws for correlators by expressing them as integrals of Schr\"odinger-covariant four-point responses weighted by the initial correlations \cite{HenkelStoimenov}. 
However, the full universal scaling functions of these correlators remain undetermined, because the four-point responses still contain functions of invariant variables that symmetry does not fix. 
This motivates a bootstrap approach that seeks further constraints from crossing symmetry and nonnegativity. 
A starting point is the decomposition of four-point responses into Schr\"odinger conformal blocks, or simply Schr\"odinger blocks \cite{Shimada}, each giving the contribution of a primary operator and its descendants, which together form a primary family.
Here we construct charged blocks in one spatial dimension for ordered four-point functions in which the two earliest insertions create two particles and the two latest annihilate them. 
In the diffusion-limited reaction $A+B\to\emptyset$ \cite{LeeCardy}, we show that nonnegativity of the unlike-species response for all positions and ordered times selects between two primary families with the same collinear function.

Schr\"odinger symmetry alone does not determine these blocks. A generic three-point function of scalar primaries contains an arbitrary function of one invariant \cite{Henkel,VolovichWen,GolkarSon}, so the dimensions and masses alone do not determine the contribution of an exchanged family to a four-point function. 
The free-boson and free-fermion four-point functions realize single charged blocks at fixed exchanged dimensions.
In the interacting Calogero model, however, the bosonic four-point function was computed exactly and shown to consist of the contribution of a single primary family in the charged channel \cite{Shimada}. Varying the coupling therefore realizes a family of charged blocks with different exchanged dimensions. 
In contrast, we here derive these blocks independently of any dynamical realization, drawing inspiration from the Casimir eigenvalue approach of Dolan and Osborn for relativistic conformal blocks \cite{DolanOsborn}.  
Specifically, we use the quartic Casimir \cite{Perroud,AizawaDobrev} and supplement its eigenvalue equation with differential constraints at the earliest and latest external insertions, which we call the endpoint equations.

We consider fundamental fields of scaling dimension $1/2$ in one spatial dimension.
In a unitary theory, a massive field of this dimension saturates the unitarity
bound and has a null Schr\"odinger descendant \cite{NishidaSonReview,Pal,Boisvert2025}. For an ordered vacuum correlator,
the corresponding vacuum condition gives homogeneous Schr\"odinger equations at
the two endpoints \cite{GolkarSon}, %
where the fields act directly on the ket and bra vacua. 
Interactions can still occur during
the middle, two-particle interval. 
We take endpoint equations as assumptions in our construction of blocks for stochastic response functions and for vacuum correlators of nonrelativistic conformal field theories. 
The broader class includes higher-dimensional systems such as fermions at unitarity, realized in ultracold atomic gases, and anyons \cite{NishidaSon,NishidaSonReview}.
In the one-dimensional setting considered here, vacuum correlators of the Calogero model satisfy these endpoint equations \cite{Shimada}.

To construct the blocks, we first describe the relative spatial positions of
the four insertions. At fixed times, removing a translation and a Galilean boost
leaves two spatial degrees of freedom, which we parametrize by signed coordinates
$(\normalA,\normalB)$. These measure departures of the adjacent triples of
insertions from straight lines in spacetime. We call a configuration collinear
when all four insertions lie on a single straight line $x=x_0+vt$. This is the
locus $\normalA=\normalB=0$. The squares of these coordinates are the three-point
invariants, but a four-point function can also depend on their relative sign.
After removing a kinematic prefactor, we work with reduced blocks $U$ that are
analytic in these coordinates and even under simultaneous spatial inversion.
Their Taylor series have an even--even sector and an odd--odd sector. The endpoint
equations determine all Taylor coefficients from two functions of the time
cross-ratio $\tau$, which we call the \emph{temporal seeds}:
\begin{equation}\label{eq:intro-seeds}
 B=U\big|_{\normalA=\normalB=0},\qquad
 Y=\tau^{-1/2}\partial_{\normalA}\partial_{\normalB}U\big|_{\normalA=\normalB=0}.
\end{equation}
Thus, within this class, we can reconstruct the full spatial dependence of $U$ from the collinear value $B$ and the rescaled mixed derivative $Y$.

We next determine the seeds $B$ and $Y$ for each exchanged primary family. 
These families are distinguished by the primary's scaling dimension $\Delta$ and by whether it is even or odd under spatial reflection. Both types of block are even under simultaneous inversion of all spatial coordinates, since the parity signs from the two three-point functions cancel.
We reduce the quartic Casimir equation to a closed system for these two functions
and prove that, together with the endpoint equations, it is equivalent to the full
Casimir problem within the class described above. 
For generic $\Delta$, the short-distance conditions imposed by the operator product expansion (OPE) then select a unique block up to normalization for each primary parity. 
This gives a derivation of the blocks independent of any dynamical
realization. 

We obtain collinear ${}_4F_3$ functions and reconstruct the full blocks as sums of products of two Kummer functions. Our construction yields both even- and odd-primary blocks for each exchanged dimension $\Delta$. The dimensions $\Delta$ and $3-\Delta$ share the same Casimir eigenvalue and form a shadow pair \cite{AizawaDobrev}. 
The Calogero model provides a concrete realization of these blocks: its bosonic and fermionic correlators correspond to the even- and odd-primary blocks, respectively. In the even-primary case, our Kummer representation is equivalent to the triple series obtained from the integral representation \cite{Shimada}.
However, the weights with which these blocks enter a general correlator remain theory-dependent.

Even the full collinear response need not fix these weights. At the one-dimensional diffusion-limited fixed point of the reaction $A+B\to\emptyset$ \cite{LeeCardy}, the unlike-species response has the same normalized collinear function for an even family of dimension two and an odd family of dimension one. 
Away from the collinear locus, in contrast, the two blocks differ. We show that nonnegativity of the unlike-species response for all positions and ordered times requires the weight of the odd family to vanish, leaving only the even family of dimension two.
The four-point response summed over particle species at each insertion also includes the same-species contribution, carried by an even family of dimension one, and thus contains the even shadow pair with dimensions $\Delta=1,2$.

The selection by nonnegativity uses the expansion in the two-particle channel,
in which the exchanged families carry a nonzero eigenvalue of the central element
$M$ of the Schr\"odinger algebra. 
Such families form the charged sector \cite{Pal},
also called the massive sector \cite{Boisvert2025}. 
Crossing symmetry requires this expansion to agree with those in crossed channels
 (Appendix~\ref{app:relabellings}), where a creation is paired with an annihilation. The exchanged operators have $M=0$ and form the neutral, or massless, sector.
For unitary nonrelativistic conformal field theories, convergence of the imaginary-time OPE has been established in certain kinematic limits \cite{GKP}.
For insertions at the same spatial point, the temporal $\mathrm{SL}(2,\mathbb R)$ subgroup gives a convergent expansion in $\hyp{2}{1}$ blocks \cite{Pal}.
These blocks describe individual $\mathrm{SL}(2,\mathbb R)$ families, whereas our collinear $\hyp{4}{3}$ functions collect the contributions of a full charged Schr\"odinger family, including spatial descendants that survive the collinear restriction.
The representation theory of the neutral sector relevant to this crossing problem has also been studied \cite{Pal,Boisvert2025,Boisvert2026}.
We use the identity contribution in the neutral OPE of the two middle insertions to normalize the charged blocks, with a corresponding sum rule for the block weights (Section~\ref{sec:checks}).
Since our derivation relies on the
nonzero exchanged mass, the full neutral-channel decompositions of these
correlators and the connection matrices relating different channel expansions
remain to be determined. A Schr\"odinger bootstrap
could use these data to constrain the block weights through crossing symmetry
and nonnegativity.

Section~\ref{sec:setting} states the assumptions and fixes the Casimir convention.
Sections~\ref{sec:kinematics}--\ref{sec:casimir} develop the signed geometry,
the endpoint lift, and the Casimir reduction to the two-function system, 
whose solutions are given in
Section~\ref{sec:solutions} as the explicit blocks. 
Section~\ref{sec:realizations} treats the Calogero model
and the diffusion-limited reaction, including the selection by nonnegativity.
Section~\ref{sec:discussion} discusses the expansion of the collinear
blocks near $\tau=1$, its potential use in a Schr\"odinger bootstrap, and further directions. 
Appendix~\ref{app:casimir} gives the explicit derivation of the Casimir identity, Appendix~\ref{app:integral} presents independent checks from the Calogero integral, and Appendix~\ref{app:invariants} examines the geometry of the four-point invariants, the kinematic relations between channels, and square-root branches.

\section{Setting and assumptions}\label{sec:setting}
\subsection{The ordered charged channel}\label{sec:ordered-channel}
Consider a vacuum four-point function at distinct ordered Euclidean times,
\begin{equation}\label{eq:correlator}
 W=\langle\Psi(t_4,x_4)\Psi(t_3,x_3)\overline\Psi(t_2,x_2)\overline\Psi(t_1,x_1)\rangle,
 \qquad t_1<t_2<t_3<t_4.
\end{equation}
The calculation uses the following assumptions.
\begin{enumerate}
 \item[(A1)] The theory has the centrally extended Schr\"odinger symmetry in one spatial
 dimension, and the vacuum and charged operator families admit the usual OPE.
 \item[(A2)] All four external fields have dimension $1/2$ and unit absolute particle mass.
 Their physical particle numbers in time order are $(+1,+1,-1,-1)$. They belong to the
 fundamental representation, which is shortened by a null descendant, and their
 earliest and latest insertions obey the homogeneous endpoint equations away from contacts.
 \item[(A3)] In the charged channel, the four-point function is in general a sum of
 contributions, each associated with a charge-two primary and its descendants.
 We isolate one such contribution, treating the scaling dimension $\Delta=2h$ of
 the exchanged primary as a parameter.
 \item[(A4)] The reduced block defined in \eqref{eq:prefactor} is analytic in the signed
 spatial variables and even under simultaneous spatial inversion. The exchanged
 primary has definite spatial parity, and the block satisfies the OPE conditions
 required by its dimension and parity
 (see Section~\ref{sec:ope-selection}).
\end{enumerate}
In a unitary Schr\"odinger theory, the fundamental representation in (A2)
saturates the unitarity bound $\delta=d/2$ \cite{NishidaSon}.
The operator corresponding to the null descendant state annihilates the vacuum.
In the ordered correlator $W$, we can use this condition at the earliest insertion
and its adjoint at the latest insertion, where the fields act directly on the
ket and bra vacua. We obtain the endpoint equations
\cite[Eqs.~(24)--(26)]{GolkarSon}:
\begin{equation}\label{eq:free-endpoints}
 (-2\partial_{t_1}-\partial_{x_1}^2)W=0,\qquad
 (2\partial_{t_4}-\partial_{x_4}^2)W=0.
\end{equation}
The condition that this operator annihilates the vacuum does not imply that it
annihilates arbitrary states.
The vacuum condition alone therefore does not imply analogous homogeneous
equations at insertions 2 and 3.
For the derivation below, we assume the endpoint equations
\eqref{eq:free-endpoints} as stated in (A2), while allowing interactions in the
two-particle interval $t_2<t<t_3$. We determine the block by combining these
equations with the Casimir eigenvalue equation for the exchanged representation.

Write $t_{ij}=t_i-t_j$ and $x_{ij}=x_i-x_j$. The time cross-ratio is
\begin{equation}\label{eq:tau}
 \tau=\frac{t_{21}t_{43}}{t_{31}t_{42}},\qquad
 1-\tau=\frac{t_{32}t_{41}}{t_{31}t_{42}},\qquad 0<\tau<1.
\end{equation}
Using the Schr\"odinger covariance (A1) \cite{Henkel,VolovichWen} and the external
dimensions and masses specified in (A2), we write the full correlator as a sum over
the charge-two families in (A3) and factor each block as
\begin{equation}\label{eq:prefactor}
 \begin{gathered}
 W=\sum_{\mathcal O_i} c_i W_{\chi_i,h_i},\qquad
 W_{\chi,h}=\cN_{\chi,h}\cP U_{\chi,h},
 \\[2pt]
 \cP=\frac1{\sqrt{t_{21}t_{43}}}
 \exp\!\left[-\frac{x_{21}^2}{2t_{21}}-\frac{x_{32}^2}{t_{32}}-\frac{x_{43}^2}{2t_{43}}\right].
 \end{gathered}
\end{equation}
The sum runs over exchanged charge-two primaries $\mathcal O_i$ of dimension
$2h_i$ and spatial parity $\chi_i$; each term includes the corresponding descendants.
Here $\cP$ is the kinematic prefactor
used in Ref.~\cite[Eq.~(3.23)]{Shimada},
and $U_{\chi,h}$ is the corresponding reduced block. The exponent is minus the sum of
the free-particle actions along the two piecewise linear worldlines $1\to2\to3$ and $2\to3\to4$,
as in \eqref{eq:B-exponent}. The factor $(t_{21}t_{43})^{-1/2}$ supplies the external
scaling weights.
The index $\chi=\pm1$ is the spatial parity of the exchanged primary.
We normalize $U_{\chi,h}$ so that its leading nonzero collinear coefficient
as $\tau\to0$ is one. The constants $\cN_{\chi,h}$ are fixed by the crossed-channel
normalization specified in Section~\ref{sec:checks}. The weights $c_i$ are
determined by the theory and the normalization of the external fields.
Until the parity $\chi$ and the dimension are specified, we write simply $U$.

\subsection{\texorpdfstring{Generators, intrinsic $\mathfrak{sl}_2$, and the Casimir eigenvalue}{Generators, intrinsic sl2, and the Casimir eigenvalue}}\label{sec:generators}
For one insertion of dimension $\delta$, on which the central element $M$
acts as the number $m$, we use the differential realization of Aizawa and Dobrev
\cite{AizawaDobrev}, %
\begin{equation}\label{eq:generators}
 \begin{aligned}
 H&=\partial_t,\quad P=\partial_x,\quad M=m,\quad G=t\partial_x+mx,\\
 D&=2t\partial_t+x\partial_x+\delta,\quad
 K=t^2\partial_t+tx\partial_x+\delta t+\frac m2x^2.
 \end{aligned}
\end{equation}
Here $m$ is the negative of the physical particle number:
$m_1=m_2=-1$ and $m_3=m_4=+1$. The early pair has $M_{12}=-2$, hence $M_{12}^2=4$.
This sign convention for insertions does not change the positive physical mass of the
propagating charge-two state.

The quartic Casimir in the same convention is \cite[Eq.~(3)]{AizawaDobrev}
\begin{equation}\label{eq:quartic}
 \widetilde C_4=(2MD-\{P,G\})^2-2\{2MK-G^2,2MH-P^2\},\quad
 \{A,B\}=AB+BA.
\end{equation}
At $M\ne0$, the generators $P$ and $G$ describe the center-of-mass motion
and, together with the central element $M$, form a Heisenberg algebra with
$[P,G]=M$. We subtract from $H$, $D$, and $K$ their center-of-mass parts,
which are quadratic in $P$ and $G$:
\begin{equation}\label{eq:subtractions}
 H_0=H-\frac{P^2}{2M},\qquad D_0=D-\frac{\{P,G\}}{2M},\qquad
 K_0=K-\frac{G^2}{2M}.
\end{equation}
These operators commute with $P,G$ and form an intrinsic $\mathfrak{sl}_2$
\cite[Theorem~5.5]{Feinsilver}.
Direct substitution gives
\begin{equation}\label{eq:omega}
 \boxed{\widetilde C_4=4M^2\Omega,\qquad
 \Omega=D_0^2-2\{K_0,H_0\}.}
\end{equation}
At fixed nonzero mass, the quartic Casimir is proportional to 
the \emph{quadratic} Casimir of the intrinsic algebra.

For a single insertion of dimension $\Delta$, set $\delta=\Delta$ in
\eqref{eq:generators}. The relation $[P,G]=m$ gives
$\{P,G\}/(2m)=GP/m+\tfrac12$, so the constant term in $D_0$ is
$d_*=\Delta-\tfrac12$. With $L=\partial_t-\partial_x^2/(2m)$,
substitution into \eqref{eq:subtractions} gives
\begin{equation}\label{eq:single-subtracted}
 H_0=L,\qquad D_0=2tL+d_*,\qquad K_0=t^2L+d_*t,
\end{equation}
and $[L,t]=1$ yields
\begin{equation}\label{eq:quadratic}
 (2tL+d_*)^2-2\{t^2L+d_*t,L\}=d_*(d_*-2).
\end{equation}
For the contribution of the charge-two family selected in (A3), the Casimir
equation is  %
\begin{equation}\label{eq:eigenvalue}
 \widetilde C_4^{(12)}W_{\chi,h}=4\Lambda_\Delta W_{\chi,h},\qquad
 \Lambda_\Delta=(2\Delta-1)(2\Delta-5)=(4h-1)(4h-5).
\end{equation}
The dimensions $\Delta$ and $3-\Delta$ are related by shadow reflection
and give the same Casimir eigenvalue \cite{AizawaDobrev}.

\section{\texorpdfstring{Four-point geometry: signed normal coordinates and invariants}{Four-point geometry: signed normal coordinates and invariants}}\label{sec:kinematics}
\subsection{The ordered real shape space}\label{sec:shape-space}
To start with, we have 8 real space-time coordinates for four ordered insertions. 
Factoring out temporal
projective transformations, translations, and boosts leaves three independent
degrees of freedom. Temporal projective
transformations fix three of the four times, leaving one temporal degree of
freedom, which we parametrize by the cross-ratio $\tau$ of \eqref{eq:tau}.
At fixed times, a translation and a Galilean boost shift each $x_i$ by
$x_0+vt_i$. Since the times are distinct, we can factor out these shifts
to obtain the two-dimensional plane
$\mathbb R^4/\operatorname{span}\{(1,1,1,1),(t_1,t_2,t_3,t_4)\}$, on which we
use the linear coordinates
\begin{equation}\label{eq:ab}
 \normalA=\frac{t_{21}x_{32}-t_{32}x_{21}}{\sqrt{2t_{21}t_{32}t_{31}}},\qquad
 \normalB=\frac{t_{32}x_{43}-t_{43}x_{32}}{\sqrt{2t_{32}t_{43}t_{42}}}.
\end{equation}
Both numerators are linear in the positions and vanish when all four insertions
lie on one straight worldline $x=x_0+vt$, so they are unchanged by these shifts.
They are also linearly independent. We take positive square roots and retain the numerator signs, as in
Ref.~\cite[Eq.~(C.6)]{Shimada} (with $\normalA=-w/\sqrt2$). Their squares are the standard three-point
invariants \cite{Henkel,VolovichWen,GolkarSon,Shimada}:
\begin{equation}\label{eq:squares}
 \normalA^2=v_{123},\qquad \normalB^2=v_{234}.
\end{equation}

We next examine how these signed coordinates transform under temporal
projective maps. Exponentiating $H$, $D$, and $K$ in \eqref{eq:generators} gives
the finite coordinate transformations $t\mapsto t'=(at+b)/(ct+d)$ and
$x\mapsto x'=x/(ct+d)$, with $ad-bc=1$ \cite{Henkel}. Such a map multiplies $\normalA$ by the sign of
$(ct_1+d)(ct_2+d)(ct_3+d)$, and $\normalB$
likewise with $t_2,t_3,t_4$. 
The two matrices
$\pm\bigl(\begin{smallmatrix}a&b\\c&d\end{smallmatrix}\bigr)$
give the same map of times and differ by the reflection $x_i\mapsto-x_i$. We restrict to maps preserving the ordering $t_1<t_2<t_3<t_4$, for which
the denominators $ct_i+d$ have the same sign at all four insertions. Of the two matrices representing the same
temporal map, we choose the one with $ct_i+d>0$, so that $\normalA$ and
$\normalB$ are invariant. 

The ordered configurations modulo these projective maps, translations,
and boosts therefore form the shape space
\begin{equation}\label{eq:shape}
 \mathcal M_4^\uparrow\simeq(0,1)_\tau\times\mathbb R^2_{\normalA,\normalB},
\end{equation}
with $(\tau,\normalA,\normalB)$ as global coordinates.
We call this parametrization by $\tau\in(0,1)$ and real
$(\normalA,\normalB)$ the \emph{ordered chart}, corresponding to the
configurations with $t_1<t_2<t_3<t_4$, with the square roots in
\eqref{eq:ab}, $\sqrt\tau$, and $\sqrt{1-\tau}$ all taken positive.

To construct an explicit representative of each point in
$\mathcal M_4^\uparrow$, we choose a frame with $t_3=0$ and
$t_4=\infty$ and use the remaining dilatation freedom to set $t_1=-1$.
With $t_2=-(1-\tau)$ and $x_3=x_4=0$, the inverse map reads
\begin{equation}\label{eq:shape-inverse}
 x_2=\sqrt{2(1-\tau)}\,\normalB,\qquad
 x_1=\sqrt{\frac2{1-\tau}}\,(\normalB+\sqrt\tau\,\normalA),
\end{equation}
and every point of $\mathcal M_4^\uparrow$ has a unique limiting representative in this frame.

The configurations on a single straight worldline form the line
$(0,1)\times\{(0,0)\}$ in $\mathcal M_4^\uparrow$, which we call the collinear locus. 
Under assumption (A4), the reduced block $U$ has a Taylor expansion
in $(\normalA,\normalB)$ about this line, with coefficients that
depend only on $\tau$. The value of $U$ and a rescaled mixed derivative
on this line define the two seed functions $B$ and $Y$ in
\eqref{eq:seeds}. The reflection
$x_i\mapsto-x_i$ acts as $(\normalA,\normalB)\mapsto(-\normalA,-\normalB)$, and
we keep it as a separate symmetry.

For three insertions the fixed-time shape is a single signed coordinate, and a
reflection-even function depends on it only through its square $v_{123}$. 
For four insertions, however,
 reflection reverses $\normalA$ and $\normalB$ together and leaves their relative
sign unchanged. 
When both $\normalA$ and $\normalB$ are nonzero, 
functions of $\normalA^2$ and $\normalB^2$ alone cannot distinguish $(\normalA,\normalB)$ from $(\normalA,-\normalB)$,
which reflection does not relate. 
We use these signed normal coordinates to retain this relative sign and to Taylor expand $U$ about the collinear locus.

\subsection{\texorpdfstring{Velocity kinks, chord residuals, and Brownian-bridge covariance}{Velocity kinks, chord residuals, and Brownian-bridge covariance}}\label{sec:residuals}
The signed normal coordinates $(\normalA,\normalB)$
are Schr\"odinger invariants with a natural
interpretation in both free nonrelativistic field theory and Brownian motion.
Within this subsection, write $p=t_{21}$, $q=t_{32}$, and $r=t_{43}$. Then
\begin{equation}\label{eq:slope}
 \normalA=\sqrt{\frac{pq}{2(p+q)}}\left(\frac{x_{32}}q-\frac{x_{21}}p\right),\qquad
 \normalA^2=\frac{x_{21}^2}{2p}+\frac{x_{32}^2}{2q}-\frac{x_{31}^2}{2(p+q)}.
\end{equation}
Thus $\normalA$ is a normalized change of segment velocity
$v_{i+1,i}=x_{i+1,i}/t_{i+1,i}$, and its square is the excess
free-particle action along the piecewise linear worldline
$1\to2\to3$ over that along the chord $13$.
For the adjacent triple $(2,3,4)$, $\normalB$ has the same interpretation.
We define the residuals of points 2 and 3 from the chords $13$ and $24$
and express the two coordinates as
\begin{equation}\label{eq:rho}
 \begin{aligned}
 \rho_{2|13}&=x_2-\frac{qx_1+px_3}{p+q},
 &\qquad \normalA&=-\frac{\rho_{2|13}}{\sqrt{2pq/(p+q)}},\\
 \rho_{3|24}&=x_3-\frac{rx_2+qx_4}{q+r},
 &\qquad \normalB&=-\frac{\rho_{3|24}}{\sqrt{2qr/(q+r)}}.
 \end{aligned}
\end{equation}

\begin{figure}[t]
 \centering
 \begin{minipage}[b]{.49\textwidth}\centering
  \includegraphics[trim=12pt 12pt 12pt 12pt,width=218.51755pt,height=187.78833pt]{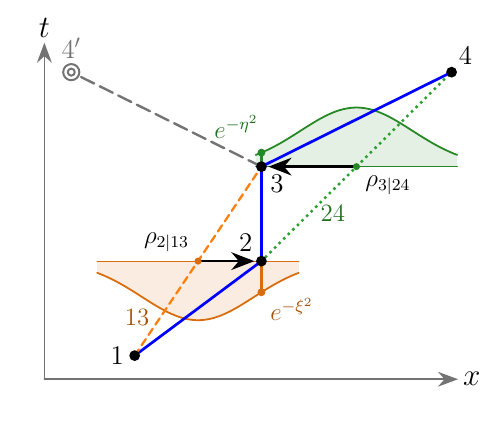}\par\vspace{-2pt}
  {\small(a) Chord residuals and bridge densities}
 \end{minipage}\hfill
 \begin{minipage}[b]{.49\textwidth}\centering
  \includegraphics[trim=12pt 12pt 12pt 12pt,width=217.66312pt,height=188.64105pt]{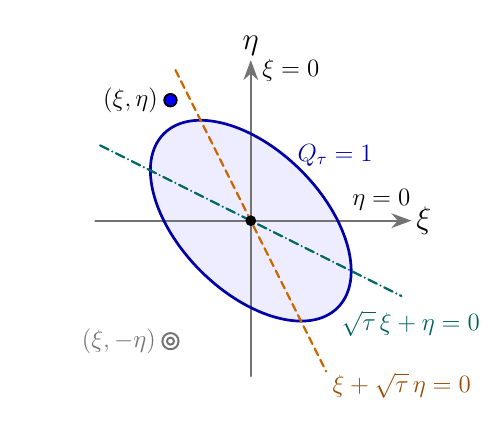}\par\vspace{-2pt}
  {\small(b) Normal plane at $\tau=0.25$}
 \end{minipage}
  \caption{(a) Four insertions in space $x$ and Euclidean time $t$, joined by a piecewise linear
 worldline (solid). The arrows from the chords $13$ (dashed) and $24$ (dotted) give the
 residuals $\rho_{2|13}$ and $\rho_{3|24}$ of \eqref{eq:rho}. The shaded profiles are the
 Brownian-bridge densities of \eqref{eq:bridge-density} at $t_2$ and $t_3$. At the insertions
 they have fallen to $e^{-\normalA^2}$ and $e^{-\normalB^2}$ of their maxima.
 Reversing the spatial displacement of $4$ from the line through $2$ and $3$ at fixed $t_4$
 gives $4'$ (double circle, long dashes). The resulting shape $(\normalA,-\normalB)$ has the
 same $v_{123}$ and $v_{234}$ but the opposite sign of $z$ in \eqref{eq:z}. The worldline
 through $4$ has an inflection ($z<0$), and the one through $4'$ has strictly monotone
 segment velocities ($z>0$). 
 The values of a block on
 the two shapes differ by $2zG$ by \eqref{eq:parity-decomposition}. 
 (b) The four triple-collinearity walls, the ellipse
 $Q_\tau=1$ of \eqref{eq:gaussian-form}, and the two shapes. 
 Here  $(\tau,\normalA,\normalB)=(0.25,-0.8,1.2)$, and (a) uses the Galilean frame with $x_2=x_3$.
 }
 \label{fig:residuals}
\end{figure}

The length scale in the denominator of \eqref{eq:rho} comes from the heat kernel
$k_\ell(x)$. This kernel solves the endpoint equations
\eqref{eq:free-endpoints} and is also the transition density of a standard Brownian
motion. Using the action difference in \eqref{eq:slope}, we express the ratio
of the heat-kernel weights for the worldline $1\to2\to3$ and the chord $13$ as
\begin{equation}\label{eq:bridge-density}
 \frac{k_{t_{21}}(x_{21})\,k_{t_{32}}(x_{32})}{k_{t_{31}}(x_{31})}
 =\frac{e^{-\normalA^2}}{\sqrt{2\pi pq/(p+q)}},
 \qquad
 k_\ell(x)=\frac{e^{-x^2/(2\ell)}}{\sqrt{2\pi\ell}}.
\end{equation}
In field-theory terms, for a free field with the unit-mass propagator
$\langle\Psi(t,x)\overline\Psi(0,0)\rangle=k_t(x)$, the left-hand side is the
three-point function $\langle\Psi(t_3,x_3)\,n(t_2,x_2)\,\overline\Psi(t_1,x_1)\rangle$
of the number density $n=\overline\Psi\Psi$, divided by
$\langle\Psi(t_3,x_3)\overline\Psi(t_1,x_1)\rangle$. It depends on the positions only
through the invariant $v_{123}=\normalA^2$ of \eqref{eq:squares}.
As a function of $x_2$, the left-hand side is the density at time $t_2$ of a Brownian
path pinned at $(t_1,x_1)$ and $(t_3,x_3)$, that is, of a Brownian bridge.
This density is Gaussian about the chord with variance $pq/(p+q)$
(see, e.g., \cite[Sec.~5.6, Eq.~(6.25)]{KaratzasShreve}). %
In this picture, $\normalA$ measures the chord residual $\rho_{2|13}$ in units of $\sqrt2$ times the
standard deviation of this bridge, and $e^{-\normalA^2}$ is the ratio of this density at $x_2$ to its
maximum on the chord (Fig.~\ref{fig:residuals}(a)).

For a standard Brownian path observed at the four insertion times, the three adjacent
increments are independent, with variances $p$, $q$, and $r$.
The residuals have zero mean, and \eqref{eq:slope} gives
$\operatorname{Var}\normalA=\operatorname{Var}\normalB=\frac12$. They share only the middle
increment $x_{32}$, with opposite signs, and $\operatorname{Var}x_{32}=q$ gives
$\operatorname{Cov}(\normalA,\normalB)=-\frac12\sqrt{pr/[(p+q)(q+r)]}=-\sqrt\tau/2$. The covariance
matrix is
\begin{equation}\label{eq:covariance}
 \Sigma_\tau=\mathbb E\!\begin{pmatrix}\normalA^2&\normalA\normalB\\\normalA\normalB&\normalB^2\end{pmatrix}
 =\frac12\begin{pmatrix}1&-\sqrt\tau\\-\sqrt\tau&1\end{pmatrix},
 \qquad \operatorname{Corr}(\normalA,\normalB)=-\sqrt\tau.
\end{equation}
Being unchanged by the boost $x_i\mapsto x_i+vt_i$, the residuals are uncorrelated with
$x_{41}$, so that the bridge pinned at insertions 1 and 4 gives them the same law.%
\footnote{The same cross-ratio appears for this bridge: its positions at $t_2$ and
$t_3$ have correlation $+\sqrt\tau$ \cite[Thm.~3 and p.~609]{HidaKuboNomotoYoshizawa}. The
sign differs because $\normalA$ and $\normalB$ are residuals sharing $x_{32}$ with opposite
signs. Appendix~\ref{app:walls} relates the two correlations by the Lorentz boost in the normal coordinates
given in \eqref{eq:signed-cross-terms}.}
The Gaussian quadratic form is
\begin{equation}\label{eq:gaussian-form}
 \begin{aligned}
 Q_\tau(\normalA,\normalB)&=\frac12(\normalA,\normalB)\Sigma_\tau^{-1}\binom{\normalA}{\normalB}
 =\frac{\normalA^2+\normalB^2+2\sqrt\tau\,\normalA\normalB}{1-\tau}
 =\frac{(\normalA+\normalB)^2}{2(1-\sqrt\tau)}+\frac{(\normalA-\normalB)^2}{2(1+\sqrt\tau)}\\
 &=\frac{x_{21}^2}{2p}+\frac{x_{32}^2}{2q}+\frac{x_{43}^2}{2r}-\frac{x_{41}^2}{2(p+q+r)}.
 \end{aligned}
\end{equation}
The last form is the excess Euclidean action of a unit-mass free particle along the
worldline $1\to2\to3\to4$ over that along the chord $14$, extending the
three-point identity for $\normalA^2$ in \eqref{eq:slope}. For this bridge, the joint density of the two intermediate positions
is proportional to $e^{-Q_\tau}$. The ellipse $Q_\tau=1$ in
Fig.~\ref{fig:residuals}(b) is therefore a contour at which this density is
$e^{-1}$ of its maximum. The same quadratic form also appears in the
free-boson block \eqref{eq:free-full} and in the hard-core boson block
\eqref{eq:B-unlike-closed} of Appendix~\ref{app:unlike}.

\subsection{The signed rational invariant \texorpdfstring{$z$}{z} and the two Taylor sectors}
Define
\begin{equation}\label{eq:z}
 z=\sqrt\tau\,\normalA\normalB=\frac{A_{123}A_{234}}{2t_{31}t_{32}t_{42}},\qquad
 A_{ijk}=\det\begin{pmatrix}1&t_i&x_i\\1&t_j&x_j\\1&t_k&x_k\end{pmatrix}.
\end{equation}
The second form shows that $z$ is rational in the original coordinates, so that it
is unaffected by consistent changes of the square-root conventions in \eqref{eq:ab},
with the sign of $\sqrt\tau$ chosen accordingly (Appendix~\ref{app:branches}). Since $\sqrt\tau>0$ on the ordered chart, \eqref{eq:slope} gives
$\operatorname{sgn}z=\operatorname{sgn}(v_{32}-v_{21})\,\operatorname{sgn}(v_{43}-v_{32})$.
Hence $z>0$ exactly when the segment velocities are strictly monotone, that is, when the
worldline bends the same way at insertions 2 and 3. Equivalently, insertions 1 and 4 lie on
the same side of the line through 2 and 3. 
For $z<0$, instead, the worldline has an inflection.

The two remaining three-point invariants are
\begin{equation}\label{eq:cross-terms}
 \begin{aligned}
 (1-\tau)v_{124}&=(\normalA+\sqrt\tau\,\normalB)^2=\normalA^2+\tau\normalB^2+2z,\\
 (1-\tau)v_{134}&=(\sqrt\tau\,\normalA+\normalB)^2=\tau\normalA^2+\normalB^2+2z,
 \end{aligned}
\end{equation}
as shown in Appendix~\ref{app:walls}, so that $2z$ is the cross term in both squares. The four
three-point invariants $v_{123}$, $v_{234}$, $v_{124}$, and $v_{134}$ vanish on the lines
$\normalA=0$, $\normalB=0$, $\normalA+\sqrt\tau\,\normalB=0$, and $\sqrt\tau\,\normalA+\normalB=0$,
respectively. These lines pass through the origin of the normal plane, which is the collinear
locus at fixed $\tau$, and form the triple-collinearity walls of Fig.~\ref{fig:residuals}(b).
Their projective cross-ratio equals the time cross-ratio $\tau$,
as shown in \eqref{eq:wall-cross-ratio}.%
\footnote{The relabelings of the insertions and the square-root branches
of the signed coordinates are collected in Appendices~\ref{app:relabellings} and~\ref{app:branches}.}

Given $\tau$, $v_{123}$, and $v_{234}$, the modulus $|z|=\sqrt{\tau v_{123}v_{234}}$ is fixed, while $v_{124}$ or $v_{134}$ determines the sign of $z$ through \eqref{eq:cross-terms}. 
Simultaneous spatial
reflection preserves $z$, whereas reversing only one signed coordinate reverses it. Two
configurations with equal $\tau$, $v_{123}$, and $v_{234}$ but opposite nonzero $z$ therefore
have different $v_{124}$ and $v_{134}$.

Consider, for example, the two shapes in Fig.~\ref{fig:residuals}.
Moving insertion $4$ alone to $4'$ reverses its spatial displacement
from the line through insertions $2$ and $3$, with all times fixed.
This changes $\normalB$ to $-\normalB$ and leaves $\normalA$ unchanged.
For $\normalA\normalB\ne0$, the shapes $(\normalA,\normalB)$ and
$(\normalA,-\normalB)$ therefore have the same $v_{123}$ and $v_{234}$ but
opposite signs of $z$. Insertions $1$ and $4$ lie on the same side of the
line through $2$ and $3$ in one shape and on opposite sides in the other,
so one shape has strictly monotone segment velocities and the other has an
inflection. Since $z$ is invariant under Schr\"odinger transformations and
overall spatial reflection, no such map relates the two shapes. Any
difference between the block values on these two shapes comes from the part
odd in $z$, which the Taylor decomposition below isolates.

In the signed coordinates, assumption (A4) requires simultaneous spatial inversion symmetry,
\begin{equation}\label{eq:U-even}
 U(\tau,-\normalA,-\normalB)=U(\tau,\normalA,\normalB).
\end{equation}
Every $U$ that is analytic in $(\normalA,\normalB)$ and obeys
\eqref{eq:U-even} decomposes uniquely, for fixed $0<\tau<1$, as
\begin{equation}\label{eq:parity-decomposition}
 U=F(\tau,\normalA^2,\normalB^2)+zG(\tau,\normalA^2,\normalB^2).
\end{equation}
Indeed, its Taylor monomials have either both powers even or both powers odd.
Since $\normalA^2=v_{123}$ and $\normalB^2=v_{234}$ by
\eqref{eq:squares}, reversing $\normalB$ alone preserves both invariants
but changes $z$ to $-z$. Hence
$U(\tau,\normalA,\normalB)-U(\tau,\normalA,-\normalB)=2zG$.
For $z\ne0$, the odd sector $zG$ measures half the difference
between the block values on the two shapes, one with strictly monotone
segment velocities and the other with an inflection.

The values of $F$ and $G$ along the collinear locus define the two seed functions
\begin{equation}\label{eq:seeds}
 B=F(\tau,0,0)=U\big|_{\normalA=\normalB=0},\qquad
 Y=G(\tau,0,0)=\tau^{-1/2}\partial_{\normalA}\partial_{\normalB}U\big|_{\normalA=\normalB=0}.
\end{equation}
This decomposition identifies the two spatial sectors. 
We show the endpoint equations determine each sector from its seed (Lemma~\ref{lem:lift}). The Casimir
equation then couples these two temporal seeds (Section~\ref{sec:seed-equivalence}), 
and the four-point problem reduces to that of two functions of $\tau$.

\section{Endpoint equations and the exact Kummer lift}\label{sec:endpoints}
\subsection{The endpoint equations in signed coordinates}
We now express \eqref{eq:free-endpoints} using the prefactor in \eqref{eq:prefactor}.
We use the frame of Section~\ref{sec:shape-space}, now retaining the scale
$T=-t_1>0$ until derivatives have been formed:
\begin{equation}\label{eq:frame}
 t_1=-T,\quad t_2=-T(1-\tau),\quad
 x_1=\sqrt{\frac{2T}{1-\tau}}(\normalB+\sqrt\tau\,\normalA),\quad
 x_2=\sqrt{2T(1-\tau)}\,\normalB,\quad T>0.
\end{equation}
The factor $\sqrt{t_4}$ in the insertion at infinity removes $t_4^{-1/2}$ from $\cP$, leaving
\begin{equation}\label{eq:prefactor-infty}
 \cP_\infty=(T\tau)^{-1/2}\exp\!\left[-2\normalB^2-
 \frac{(\normalA+\sqrt\tau\,\normalB)^2}{1-\tau}\right].
\end{equation}
Define $\theta_T=T\partial_T$ at fixed $(\tau,\normalA,\normalB)$ and $\theta=\tau\partial_\tau$ at fixed $(T,\normalA,\normalB)$.
The prefactor $\cP_\infty$ accounts for the scaling weight in this frame,
so dilatation invariance gives $\theta_T U=0$ at fixed $(\tau,\normalA,\normalB)$.
For the earliest endpoint, conjugation by $\cP_\infty$ and the chain rule give
\begin{equation}\label{eq:chain-endpoint}
 \cP_\infty^{-1}(-2\partial_{t_1}-\partial_{x_1}^2)\cP_\infty
 =\frac1T\left[2\theta_T-\frac{1-\tau}{2\tau}
 (\partial_{\normalA}^2-2\normalA\partial_{\normalA}-4\theta)\right].
\end{equation}
The derivative rules are given in \eqref{eq:A-derivatives} and \eqref{eq:A-chain}.
We obtain the corresponding conjugated operator for the latest endpoint by time reversal
and charge conjugation, with the relabeling $t'_i=-t_{5-i}$, $x'_i=x_{5-i}$. The relabeling preserves $\tau$ and $\cP$ and interchanges
$\normalA$ and $\normalB$.
With $\theta_T U=0$, the two endpoint equations \eqref{eq:free-endpoints},
which are homogeneous Schr\"odinger equations for $W$,
take the Ornstein--Uhlenbeck form for $U$:
\begin{equation}\label{eq:endpoints}
 \boxed{L_{\normalA}U=L_{\normalB}U=0,\quad
 L_{\normalA}=\partial_{\normalA}^2-2\normalA\partial_{\normalA}-4\theta,\quad
 L_{\normalB}=\partial_{\normalB}^2-2\normalB\partial_{\normalB}-4\theta.}
\end{equation}
As an aside, we note that setting
$\psi=e^{-(\normalA^2+\normalB^2)/2}U$
puts the endpoint equations in harmonic-oscillator form:
\begin{equation}\label{eq:oscillator}
 (-\partial_{\normalA}^2+\normalA^2+4\theta-1)\psi=0,\qquad
 (-\partial_{\normalB}^2+\normalB^2+4\theta-1)\psi=0.
\end{equation}
After continuation to imaginary $(\normalA,\normalB)$, the oscillator
eigenfunctions provide a way to organize descendant contributions,
as illustrated by the second term of the free-boson block
\eqref{eq:free-full}. %

\subsection{Two seeds determine every spatial coefficient}
The seeds \eqref{eq:seeds} are the collinear value $B$ and the rescaled mixed derivative $Y$:
\begin{equation}\label{eq:jets}
 B=U\big|_{\normalA=\normalB=0},\qquad
 J=\partial_{\normalA}\partial_{\normalB}U\big|_{\normalA=\normalB=0},\qquad Y=\tau^{-1/2}J.
\end{equation}
Thus $Y$ is the mixed derivative $J$ with its temporal power shifted by $-1/2$.
The first terms are
\begin{equation}\label{eq:jet-expansion}
 U=B+2\theta B(\normalA^2+\normalB^2)+zY+O\bigl((\normalA,\normalB)^4\bigr).
\end{equation}
For $U_{ij}=\partial_{\normalA}^i\partial_{\normalB}^jU|_{\normalA=\normalB=0}$,
the endpoint equations \eqref{eq:endpoints} imply
\begin{equation}\label{eq:taylor-rec}
 U_{i+2,j}=(4\theta+2i)U_{ij},\qquad
 U_{i,j+2}=(4\theta+2j)U_{ij}.
\end{equation}
Condition \eqref{eq:U-even} removes $U_{10},U_{01}$, leaving the seeds $U_{00}=B$ and $U_{11}=J$.
In particular,
\begin{equation}\label{eq:jet-derivatives}
 U_{20}=U_{02}=4\theta B,\qquad U_{22}=16\theta^2B.
\end{equation}
The Casimir operator also probes the mixed derivative $J$, and the identity for $U_{22}$
is used in Section~\ref{sec:seed-equivalence} to obtain the coupled equations for $B$ and $Y$.
Once $B$ and $Y$ are determined, one can reconstruct the full block $U$ using the
endpoint recurrences, as the following lemma shows.\footnote{%
We call this reconstruction the \emph{Kummer lift}.
It maps the collinear seeds $(B,Y)$ to a formal spatial solution
of the two endpoint equations using Kummer's confluent hypergeometric
function ${}_1F_1$.
The term ``Kummer lift'' is also used in number theory
with a different meaning.}

\begin{lemma}[Exact Kummer endpoint lift]\label{lem:lift}
Fix a branch of $\sqrt\tau$ away from $\tau=0$ (see Appendix~\ref{app:branches}).
The map $U\mapsto(B,Y)$ defined by \eqref{eq:jets} is a bijection between formal spatial
power-series solutions of \eqref{eq:endpoints} satisfying \eqref{eq:U-even} and pairs of smooth temporal seed functions. Its inverse is
\begin{equation}\label{eq:lift}
 \begin{aligned}
 \operatorname{Lift}(B,Y)={}&
 \hyp{1}{1}(\theta;\tfrac12;\normalA^2)
 \hyp{1}{1}(\theta;\tfrac12;\normalB^2)B\\
 &+z\hyp{1}{1}(\theta+1;\tfrac32;\normalA^2)
 \hyp{1}{1}(\theta+1;\tfrac32;\normalB^2)Y.
 \end{aligned}
\end{equation}
Here the Kummer series are understood formally, with $\theta$ acting on $B$
in the first line and on $Y$ in the second. Any such solution that is analytic in
$(\normalA,\normalB)$ near $(0,0)$ is uniquely determined by its seeds $(B,Y)$.
\end{lemma}
\begin{proof}
Iterating \eqref{eq:taylor-rec} fixes all even--even coefficients from $B$ and all
odd--odd coefficients from $J$. The two recursion directions commute, since their
coefficients are polynomials in $\theta$. The coefficient of $\normalA^{2m}\normalB^{2n}$ is
$4^{m+n}(\theta)_m(\theta)_nB/[(2m)!(2n)!]$. Using
$(2m)!=4^m m!(1/2)_m$ gives the first line of \eqref{eq:lift}.
For the odd--odd sector, $J=\sqrt\tau Y$ and
$(\theta+1/2)_m\sqrt\tau=\sqrt\tau(\theta+1)_m$ give the second line.
Conversely these coefficients obey both recurrences and have the stated seeds.
\end{proof}
It is indeed possible to obtain analytic lifts from these formal series, as we show in Appendix~\ref{app:gaussian-moments} by proving convergence throughout the shape space $\mathcal M_4^\uparrow$ for the even- and odd-primary branches of Section~\ref{sec:solutions} with half-dimension $q\ge1/2$.
For other generic $q$, convergence of the Kummer sums follows from
Appendix~\ref{app:branches}.

For a temporal power $\tau^\lambda$, the one-variable equation and its normalized even and
odd solutions are
\begin{align}
 f''(\normalA)-2\normalA f'(\normalA)-4\lambda f(\normalA)&=0,\label{eq:kummer-ode}\\
 E_\lambda(\normalA)=\hyp{1}{1}(\lambda;\tfrac12;\normalA^2),\qquad
 O_\lambda(\normalA)&=\normalA\hyp{1}{1}(\lambda+\tfrac12;\tfrac32;\normalA^2).
 \label{eq:even-odd}
\end{align}
Putting $f(\normalA)=g(\normalA^2)$ or $f(\normalA)=\normalA g(\normalA^2)$ gives the corresponding Kummer equations directly. 
For later use, we note the derivative identities
\begin{equation}\label{eq:kummer-derivatives}
 E_\lambda'(\normalA)=4\lambda O_{\lambda+1/2}(\normalA),\qquad O_\lambda'(\normalA)=E_{\lambda+1/2}(\normalA).
\end{equation}

\section{Casimir reduction and the coupled temporal system}\label{sec:casimir}
\subsection{The exact operator identity}
For every generator in \eqref{eq:quartic}, take its pair sum $X_{12}=X_1+X_2$.
We introduce the \emph{primary--shadow polynomial}
\begin{equation}\label{eq:AhDh}
 p_h(\theta)=(\theta-h)(\theta+h-\tfrac32),
\end{equation}
whose roots $h$ and $3/2-h$ are the half-dimensions of a primary and its shadow.
On a Schr\"odinger-covariant function $\cP U(\tau,\normalA,\normalB)$ the following identity is exact,
before imposing either endpoint equation:
\begin{equation}\label{eq:operator-identity}
 \begin{aligned}
 \frac1{16}\cP^{-1}(\widetilde C_4^{(12)}-4\Lambda_\Delta)\cP
 &=4p_h(\theta)-\sqrt\tau(\theta-\tfrac12)\partial_{\normalA}\partial_{\normalB}\\
 &\quad+\frac\tau4\bigl[(2\theta+\normalB\partial_{\normalB})L_{\normalA}
 +(2\theta+\normalA\partial_{\normalA})L_{\normalB}\bigr].
 \end{aligned}
\end{equation}
Here each operator acts on all factors to its right. In particular, $\theta$ differentiates the
$\tau$ in the endpoint operators, whereas the leftmost factor $\tau/4$ stays on the
left. Appendix~\ref{app:casimir} derives this identity from the original generators and
records the complete intermediate expression.

When $U$ satisfies the two endpoint equations, the last line of
\eqref{eq:operator-identity} vanishes upon acting on $U$, and the full Casimir equation becomes
\begin{equation}\label{eq:reduced}
 \boxed{4p_h(\theta)U=\sqrt\tau(\theta-\tfrac12)
 \partial_{\normalA}\partial_{\normalB}U.}
\end{equation}
On the collinear locus, a term proportional to $\normalA\normalB$ contributes to
\eqref{eq:reduced} through its mixed derivative, even though the term itself vanishes.

\subsection{\texorpdfstring{Equivalence of the seed equations and the full Casimir equation}{Equivalence of the seed equations and the full Casimir equation}}\label{sec:seed-equivalence}
Evaluating \eqref{eq:reduced} on the collinear locus $\normalA=\normalB=0$,
with the spatial derivatives taken before restriction, gives the first equation below,
while applying an additional mixed derivative $\partial_\normalA\partial_\normalB$
to \eqref{eq:reduced} before restriction and using $U_{22}=16\theta^2B$
from \eqref{eq:jet-derivatives} gives the second:
\begin{align}
 4p_h(\theta)B&=\sqrt\tau(\theta-\tfrac12)J=\tau\theta Y,\label{eq:BJ}\\
 4p_h(\theta)J&=16\sqrt\tau\,\theta^2(\theta-\tfrac12)B.\label{eq:JB}
\end{align}
Since $p_h(\theta)\sqrt\tau=\sqrt\tau p_h(\theta+\tfrac12)$, the result is
\begin{equation}\label{eq:BY}
 \boxed{\begin{aligned}
 4(\theta-h)(\theta+h-\tfrac32)B&=\tau\theta Y,\\
 (\theta-h+\tfrac12)(\theta+h-1)Y&=4\theta^2(\theta-\tfrac12)B.
 \end{aligned}}
\end{equation}
Equivalently,
\begin{equation}\label{eq:seed-matrix}
 \mathcal C_h(\tau,\theta)\binom BY=0,\qquad
 \mathcal C_h=\begin{pmatrix}
 4p_h(\theta)&-\tau\theta\\
 -4\theta^2(\theta-\tfrac12)&p_h(\theta+\tfrac12)
 \end{pmatrix}.
\end{equation}
We have introduced the rescaled seed $Y=\tau^{-1/2}J$ to remove the explicit
$\sqrt\tau$ factors from \eqref{eq:BJ}--\eqref{eq:JB}.
Both equations for $(B,J)$ contain the same primary--shadow polynomial
$p_h(\theta)$, while after rescaling, the second diagonal entry of
$\mathcal C_h$ contains $p_h(\theta+\tfrac12)$, with the shift matching
the temporal weight of $z=\sqrt\tau\,\normalA\normalB$.
With this choice, the Frobenius series for $B$ and $Y$
selected by the single-family OPE condition start at the same power
for an even primary and at powers differing by one for an odd primary,
as explained in Section~\ref{sec:ope-selection}.

In order to prove that the two seed equations are sufficient, we set
\begin{equation}\label{eq:Th}
 T_h=4p_h(\theta)-\sqrt\tau(\theta-\tfrac12)\partial_{\normalA}\partial_{\normalB}.
\end{equation}
Since the following elementary commutators
\begin{equation}\label{eq:commutators}
 [L_{\normalA},\theta]=0,\quad [L_{\normalA},\sqrt\tau]=-2\sqrt\tau,\quad
 [L_{\normalA},\partial_{\normalA}]=2\partial_{\normalA},\quad
 [L_{\normalA},\partial_{\normalB}]=0
\end{equation}
give $[L_{\normalA},T_h]=[L_{\normalB},T_h]=0$, $T_hU$ itself satisfies the same endpoint
recurrences as $U$ and inherits its simultaneous spatial inversion symmetry
\eqref{eq:U-even}. Since its collinear value and rescaled mixed derivative
vanish by \eqref{eq:BY}, the Kummer lift of Lemma~\ref{lem:lift}
reconstructs $T_hU$ from the zero seeds $(B,Y)=(0,0)$, giving $T_hU=0$.
This is precisely \eqref{eq:reduced}, to which the full Casimir equation reduces
when the two endpoint equations hold. %
Together with the necessity established above, this
proves that the seed equations \eqref{eq:BY} are equivalent to the full Casimir
equation for analytic solutions of the two endpoint equations with simultaneous
spatial inversion symmetry \eqref{eq:U-even}.

\subsection{\texorpdfstring
  {The fourth-order equation for $B$ and its local exponents}
  {The fourth-order equation for B and its local exponents}}
For any polynomial $f$, $f(\theta-1)\tau=\tau f(\theta)$. Apply $p_h(\theta-\tfrac12)$ to
the first equation in \eqref{eq:BY} and use the second. 
This eliminates $Y$ and gives the fourth-order scalar equation for $B$:
\begin{equation}\label{eq:fourth-order}
 \bigl[(\theta-h)(\theta-h-\tfrac12)(\theta+h-\tfrac32)(\theta+h-2)
 -\tau\theta^3(\theta-\tfrac12)\bigr]B=0.
\end{equation}
Substituting $B\sim\tau^q$ into \eqref{eq:fourth-order} gives the indicial
polynomial $p_h(q)p_h(q-\tfrac12)$ at $\tau=0$, so the four generic
exponents are
\begin{equation}\label{eq:four-exponents}
 h,\qquad h+\tfrac12,\qquad \tfrac32-h,\qquad 2-h.
\end{equation}
These exponents correspond to the two primary parities at each of the two
dimensions $\Delta$ and $3-\Delta$, as established next. 
For $0<\tau<1$, the coupled equations for $(B,Y)$ can be written
as a system of four first-order differential equations, whose solution space is four-dimensional, like that of
\eqref{eq:fourth-order}. At $h=1/2$ or $h=1$, however, one can find a nonzero
solution $(B,Y)$ with $B(\tau)\equiv0$, as shown explicitly in \eqref{eq:invisible-seed},
so that the collinear value does not determine the full block. We
therefore construct the full blocks from the coupled equations rather
than from the scalar equation \eqref{eq:fourth-order} alone.

\section{\texorpdfstring{Explicit blocks: four local branches}{Explicit blocks: four local branches}}\label{sec:solutions}
\subsection{OPE branch selection from descendants}\label{sec:ope-selection}
Let $q$ denote the half-dimension of the chosen branch, so
$q\in\{h,3/2-h\}$ at fixed Casimir. In the state convention, where $P$ and $H$ raise
dimension \cite{NishidaSon}, a charged family has descendants
\begin{equation}\label{eq:descendants}
 |n,m\rangle=P^nH_0^m|\mathcal O_{\chi}\rangle,\qquad
 \Delta_{n,m}=2q+n+2m,\qquad n,m\ge0.
\end{equation}
Here $H_0$ is the subtracted generator of \eqref{eq:subtractions} acting
on states, and it commutes with $P$. Spatial inversion changes the sign of $P$ and preserves $H_0$.
If the exchanged primary has spatial parity $\chi$, its descendant
$P^nH_0^m|\mathcal O_\chi\rangle$ has parity $(-1)^n\chi$.
At collinearity, one can use a translation and a boost to put all four external
insertions at $x=0$.
Each pair of external scalar fields is then even under spatial
inversion and couples only to spatially even intermediate states.
    Thus
\begin{equation}\label{eq:first-states}
 \begin{array}{c|c|c}
 \text{primary parity}&\text{first state visible in }B&\text{dimension}\\\hline
 +1&|\mathcal O_+\rangle&2q\\
 -1&P|\mathcal O_-\rangle&2q+1
 \end{array}
\end{equation}
Since time has scaling weight two, these dimensions give leading collinear powers
$\tau^q$ and $\tau^{q+1/2}$. Subsequent allowed values of $n$ differ by two and
$H_0$ adds two, giving integer steps in $\tau$.

The label $\chi$ is the primary parity under spatial inversion.
In the two-body realization, exchange reverses the relative coordinate,
and the relative wave function is \emph{extended} from one side of contact to
the other either symmetrically or antisymmetrically under exchange.
Since $P$ differentiates only the center of mass, it preserves this
exchange extension. For a primary with an even center-of-mass ground state, its spatial parity
agrees with the exchange parity of the relative wave function, giving the
bosonic and fermionic realizations of Section~\ref{sec:realizations}.
Note that blocks of either primary parity $\chi$ are invariant 
under simultaneous spatial inversion as in \eqref{eq:U-even}.

For the even primary, using this counting, we write %
\begin{equation}\label{eq:ope}
 B_{+,q}=\tau^q\sum_{n=0}^\infty b_n^{(+)}\tau^n,\qquad
 Y_{+,q}=\tau^q\sum_{n=0}^\infty y_n^{(+)}\tau^n,\qquad b_0^{(+)}=1,
\end{equation}
where the superscript $(\pm)$ on the coefficients denotes the primary parity.
We have imposed the single-family OPE condition of (A4) by requiring
$B_{+,q}/\tau^q$ and $Y_{+,q}/\tau^q$ to be power series in $\tau$,
a condition stronger than simply $B\sim\tau^q$.
The corresponding expansion of the unrescaled mixed seed $J=\sqrt\tau\,Y$
begins a half-step higher at $\tau^{q+1/2}$.
For the odd primary, $J$ instead starts at $\tau^q$,
while $B$ first appears at $\tau^{q+1/2}$.

\subsection{The even-primary branch}\label{sec:even}
For either choice $q=h$ or $q=3/2-h$, the primary--shadow polynomial
satisfies $p_q(\theta)=p_h(\theta)$.
Substituting \eqref{eq:ope} into \eqref{eq:BY} gives:
\begin{equation}\label{eq:coeff-rec}
 \begin{aligned}
 4n(2q+n-\tfrac32)b_n^{(+)}&=(q+n-1)y_{n-1}^{(+)},&&n\ge1,\\
 (n+\tfrac12)(2q+n-1)y_n^{(+)}&=4(q+n)^2(q+n-\tfrac12)b_n^{(+)},&&n\ge0.
 \end{aligned}
\end{equation}
Generically $y_0^{(+)}=4q^2$. Eliminating $y_{n-1}^{(+)}$ gives
\begin{equation}\label{eq:ratio}
 \frac{b_n^{(+)}}{b_{n-1}^{(+)}}=\frac{(q+n-1)^3(q+n-\tfrac32)}
 {n(n-\tfrac12)(2q+n-\tfrac32)(2q+n-2)}.
\end{equation}
With $(c)_n=c(c+1)\cdots(c+n-1)$, the two coefficient sequences are
\begin{equation}\label{eq:coefficients}
 b_n^{(+)}(q)=\frac{(q)_n^3(q-\tfrac12)_n}
 {n!(\tfrac12)_n(2q-\tfrac12)_n(2q-1)_n},\qquad
 y_n^{(+)}(q)=4q^2\frac{(q+1)_n^2(q)_n(q+\tfrac12)_n}
 {n!(\tfrac32)_n(2q-\tfrac12)_n(2q)_n}.
\end{equation}
The collinear solution follows immediately:
\begin{equation}\label{eq:collinear}
 \boxed{B_{+,q}(\tau)=\tau^q\hyp{4}{3}\!\left(
 \begin{matrix}q,q,q,q-\tfrac12\\\tfrac12,2q-\tfrac12,2q-1\end{matrix};\tau\right).}
\end{equation}
We first use these formulas for values of $q$ at which none of the lower
parameters is zero or a negative integer. 
At the removable resonances $q=1/2,0,-1/2,-1,\ldots$,
we take the limit as $q$ approaches
the resonant value, substituting the displayed expressions for all
upper and lower parameters before taking the limit.
The remaining parameter poles occur at $q=1/4,-1/4,-3/4,\ldots$,
where subtracting the pole terms using the shadow branches 
(Section~\ref{sec:shadow}) gives finite limits containing $\log\tau$.

Apply Lemma~\ref{lem:lift} term by term to reconstruct the full block:
\begin{equation}\label{eq:full-block}
 \begin{aligned}
 U_{+,q}(\tau,\normalA,\normalB)={}&\sum_{n\ge0}b_n^{(+)}(q)\tau^{q+n}
 E_{q+n}(\normalA)E_{q+n}(\normalB)\\
 &+\sum_{n\ge0}y_n^{(+)}(q)\tau^{q+n+1/2}
 O_{q+n+1/2}(\normalA)O_{q+n+1/2}(\normalB).
 \end{aligned}
\end{equation}
We see the derivative identities \eqref{eq:kummer-derivatives} 
reduce the full equation \eqref{eq:reduced} to exactly \eqref{eq:coeff-rec}, 
providing a direct check of the lift.

To express the block as a triple series, define
$\cF_{+,q}=\tau^{-q}U_{+,q}$.
Here $z=\sqrt\tau\,\normalA\normalB$ is the signed invariant
defined in \eqref{eq:z}, with $z^2=\tau v_{123}v_{234}$.
Expanding the Kummer functions gives the triple series obtained
for the bosonic Calogero four-point function in
Ref.~\cite[Eqs.~(3.24)--(3.25)]{Shimada}
(see \eqref{eq:calogero-identification} for the parameter identification):
\begin{equation}\label{eq:triple}
 \cF_{+,q}=\sum_{p,m,n\ge0}\frac{\tau^p v_{123}^m v_{234}^n}{p!m!n!}
 \bigl[\Lambda_{pmn}^{(0)}+z\Lambda_{pmn}^{(1)}\bigr],
\end{equation}
where
\begin{equation}\label{eq:lambda-coeff}
 \Lambda_{pmn}^{(j)}=(2q)^{2j}
 \frac{(q+j)_{p+m}(q+j)_{p+n}(2q-1+j)_{2p}}
 {(\tfrac12+j)_m(\tfrac12+j)_n(\tfrac12+j)_p(4q-2+j)_{2p}},\quad j=0,1.
\end{equation}
Only the product identities
\begin{equation}\label{eq:pochhammer}
 (c)_{p+m}=(c)_p(c+p)_m,\qquad
 (c)_{2p}=2^{2p}(c/2)_p\bigl((c+1)/2\bigr)_p
\end{equation}
are needed to obtain \eqref{eq:lambda-coeff} from \eqref{eq:full-block}
using the coefficients in \eqref{eq:coefficients}.
The square-root factor $\sqrt{\tau v_{123}v_{234}}$ in that
reference should be read as the signed invariant $z$.
Since $\Lambda^{(1)}_{000}=4q^2$, the series contains a nonzero
term linear in $z$ for $q\ne0$, and the block depends on the sign
of $z$ even at fixed $(\tau,v_{123},v_{234})$.\footnote{For the two shapes of
Fig.~\ref{fig:residuals}, $(\tau,\normalA,\normalB)=(0.25,-0.8,\pm1.2)$, the Kummer sum
\eqref{eq:full-block} at $q=5/4$ gives $\cF_{+,5/4}\simeq21.5$ for the inflected shape
($z=-0.48$) and $\cF_{+,5/4}\simeq384$ for the monotone one ($z=0.48$).}

Appendix~\ref{app:gaussian-moments} also proves convergence throughout
the shape space $\mathcal M_4^\uparrow$ for $q\ge1/2$;
for other generic $q$, see Appendix~\ref{app:branches}.

\subsection{The odd-primary branch}\label{sec:odd}
Using the odd-primary OPE powers described in
Section~\ref{sec:ope-selection} and $Y=\tau^{-1/2}J$,
we normalize the first nonzero collinear coefficient to one and write
\begin{equation}\label{eq:odd-seeds}
 B_{-,q}=\tau^{q+1/2}\sum_{n\ge0}b_n^{(-)}\tau^n,\qquad
 Y_{-,q}=4\tau^{q-1/2}\sum_{n\ge0}y_n^{(-)}\tau^n,\qquad
 b_0^{(-)}=y_0^{(-)}=1.
\end{equation}
For generic $q$, the factor four follows at leading order from
the first equation of \eqref{eq:BY}.
At $q=1/2$, the recursions leave $y_0^{(+)}$ and $y_0^{(-)}$ free.
We define both branches there by the parameter limit described above
for removable resonances, which selects $y_0^{(+)}=y_0^{(-)}=1$.
Substitution into that same system gives
\begin{equation}\label{eq:odd-rec}
 \begin{aligned}
 (n+\tfrac12)(2q+n-1)b_n^{(-)}
 &=(q+n-\tfrac12)y_n^{(-)},&&n\ge0,\\
 n(2q+n-\tfrac32)y_n^{(-)}
 &=(q+n-\tfrac12)^2(q+n-1)b_{n-1}^{(-)},&&n\ge1.
 \end{aligned}
\end{equation}
Consequently
\begin{equation}\label{eq:odd-coefficients}
 b_n^{(-)}(q)=\frac{(q+\tfrac12)_n^3(q)_n}
 {n!(\tfrac32)_n(2q-\tfrac12)_n(2q)_n},\qquad
 y_n^{(-)}(q)=\frac{(q+\tfrac12)_n^2(q)_n(q-\tfrac12)_n}
 {n!(\tfrac12)_n(2q-\tfrac12)_n(2q-1)_n}.
\end{equation}
The complete odd block and its collinear value are
\begin{equation}\label{eq:odd-full}
 \begin{aligned}
 U_{-,q}={}&4\sum_{n\ge0}y_n^{(-)}(q)\tau^{q+n}
 O_{q+n}(\normalA)O_{q+n}(\normalB)\\
 &+\sum_{n\ge0}b_n^{(-)}(q)\tau^{q+n+1/2}
 E_{q+n+1/2}(\normalA)E_{q+n+1/2}(\normalB),
 \end{aligned}
\end{equation}
\begin{equation}\label{eq:odd-collinear}
 \boxed{B_{-,q}(\tau)=\tau^{q+1/2}\hyp{4}{3}\!\left(
 \begin{matrix}q+\tfrac12,q+\tfrac12,q+\tfrac12,q\\\tfrac32,2q-\tfrac12,2q\end{matrix};\tau\right).}
\end{equation}
The leading term of the full block is
$4\tau^qO_q(\normalA)O_q(\normalB)$, which vanishes on the collinear locus.
There the first visible state is the next spatial level $P|\mathcal O_-\rangle$, as in
\eqref{eq:first-states}. The collinear value $B_{-,q}$ has the same leading power and the
same upper parameters as $B_{+,q+1/2}$ in \eqref{eq:collinear}, while its lower parameters
$(\tfrac32,2q-\tfrac12,2q)$ replace $(\tfrac12,2q+\tfrac12,2q)$.
The corresponding full blocks have primary dimensions $2q$ and $2q+1$,
and hence Casimir eigenvalues $\Lambda_{2q}$ and $\Lambda_{2q+1}$, respectively.

The primary parity therefore fixes the spatial sector of the leading
OPE term only. For $q\ne0$, the terms with coefficients $y_n^{(+)}$
in \eqref{eq:full-block} and $b_n^{(-)}$ in \eqref{eq:odd-full}
also contribute at subleading orders to the odd--odd and even--even
spatial sectors, respectively, so both branches use both seeds.

\subsection{The shadow branches}\label{sec:shadow}
The primary--shadow polynomial is invariant under $h\mapsto3/2-h$:
$p_{3/2-h}(\theta)=p_h(\theta)$.
Writing $\nu=2h-3/2$, the Casimir eigenvalue fixes $\nu^2$, and the shadow reflection is
\begin{equation}\label{eq:shadow}
 (\Delta,M,\chi)\longmapsto(3-\Delta,M,\chi),\qquad \nu\longmapsto-\nu.
\end{equation}
It changes the scaling branch and preserves both the charge $M$
and the primary parity $\chi$. The number 3 comes from the
scaling weight of $dt\,dx$; intertwining operators relate representations
of weights $\Delta$ and $3-\Delta$
\cite[Sec.~5]{AizawaDobrev}. The four generic local branches are summarized below,
with the leading seed terms obtained from \eqref{eq:coefficients} and \eqref{eq:odd-seeds}.
\begin{center}
 \begin{tabular}{@{}llll@{}}\toprule
 Primary parity & Half-dimension & Leading $B$ & Leading $Y$\\\midrule
 Even & $h$ & $\tau^h$ & $4h^2\tau^h$\\
 Even & $\widetilde h=3/2-h$ & $\tau^{\widetilde h}$ & $4\widetilde h^2\tau^{\widetilde h}$\\
 \addlinespace
 Odd & $h$ & $\tau^{h+1/2}$ & $4\tau^{h-1/2}$\\
 Odd & $\widetilde h=3/2-h$ & $\tau^{\widetilde h+1/2}$ & $4\tau^{\widetilde h-1/2}$\\\bottomrule
 \end{tabular}
\end{center}
We observe these reproduce \eqref{eq:four-exponents}. 
The four blocks form a local basis for generic $h$. At $h=3/4$ ($\Lambda=-4$),
the shadow branches coincide, and a complete local basis requires logarithmic solutions.
At $\Lambda=-3$, a different phenomenon occurs, where distinct full blocks have the same
collinear value, as shown below.

\subsection{\texorpdfstring{A solution invisible to the collinear value at $\Lambda=-3$}{A solution invisible to the collinear value at Lambda = -3}}\label{sec:resonance}
Setting $B=0$ in \eqref{eq:BY}, one has $\theta Y=0$ and
$p_h(\tfrac12)Y=(1/2-h)(h-1)Y=0$, which means that a nonzero $Y$ is possible only
for $h=1/2$ or $h=1$, corresponding to $\Lambda=-3$. For either value of $h$,
\begin{equation}\label{eq:invisible-seed}
 (B,Y)=(0,1)
\end{equation}
is a nonzero solution. The Kummer endpoint lift of this seed pair to the full block yields %
\begin{equation}\label{eq:invisible-lift}
 U_{\mathrm{inv}}=z\hyp{1}{1}(1;\tfrac32;\normalA^2)
 \hyp{1}{1}(1;\tfrac32;\normalB^2).
\end{equation}
It obeys both endpoint equations and the full Casimir equation, while vanishing on
the entire collinear locus $\normalA=\normalB=0$.
Taking $q\to1/2$ in the odd branch and setting $q=1$ in the even branch, we obtain
\begin{equation}\label{eq:resonance-difference}
 B_{-,1/2}=B_{+,1},\qquad Y_{-,1/2}-Y_{+,1}=4,
 \qquad U_{-,1/2}-U_{+,1}=4U_{\mathrm{inv}}.
\end{equation}
Comparing coefficients of equal powers of $\tau$ in \eqref{eq:coefficients}
and \eqref{eq:odd-coefficients} gives the same $B$ series and a constant difference
of $4$ between the $Y$ series. The identity for the full blocks then follows from
the linearity of the lift. Adding a multiple of $U_{\mathrm{inv}}$ leaves
the collinear value unchanged.

The four solutions $U_{+,1/2}$, $U_{-,1}$, $U_{+,1}$, and $U_{\mathrm{inv}}$
form a local basis at $\Lambda=-3$.\footnote{The scalar equation
\eqref{eq:fourth-order} has a fourth local solution containing $\tau\log\tau$
in addition to the three independent collinear values of this basis.
For this logarithmic $B$, the first equation in \eqref{eq:BY} determines $Y$
up to an additive constant. The two sides of the second equation then differ by
a nonzero constant. Since $p_h(\tfrac12)=0$, changing the additive constant in $Y$
leaves this difference unchanged. Therefore no $Y$ completes this logarithmic $B$
to a solution of \eqref{eq:BY}, and no full block has this collinear value.}
To determine which combination occurs in a physical correlator,
we need further input, such as the nonnegativity used in
Section~\ref{sec:dla-signed}.

\subsection{Elementary blocks, normalization, and a crossed-channel sum rule}\label{sec:checks}
At the free-boson value $q=1/2$, an upper and a lower parameter in
\eqref{eq:collinear} vanish together. Taking their joint parameter limit gives
\begin{equation}\label{eq:free-collinear}
 B_{+,1/2}(\tau)=\frac{\sqrt\tau}{2}
 \left(1+\frac1{\sqrt{1-\tau}}\right).
\end{equation}
The full block, in terms of $\cF_{+,q}=\tau^{-q}U_{+,q}$, is
\begin{equation}\label{eq:free-full}
 \cF_{+,1/2}(\tau,\normalA,\normalB)=\frac12e^{\normalA^2+\normalB^2}
 +\frac1{2\sqrt{1-\tau}}\exp\!\left(
 \frac{\normalA^2+\normalB^2+2\sqrt\tau\,\normalA\normalB}{1-\tau}\right).
\end{equation}
Appendix~\ref{app:gaussian-moments} obtains this from two Gaussian integrals. 
The second term has a Mehler form\footnote{After analytic continuation to
$(\normalA,\normalB)=(ia,ib)$ with real $a,b$, this term is 
$\tfrac12 e^{-a^2-b^2}$ times the Hermite series
in Mehler's formula with parameter $-\sqrt\tau$
\cite[Eq.~(18.18.28)]{DLMF}.
With the Gaussian factor used in \eqref{eq:oscillator}, this gives an
oscillator eigenfunction expansion of the corresponding contribution to
the analytically continued $\psi$,
reorganizing the descendant contributions to this term into oscillator modes.
A related sum over the relative-motion descendants of the Calogero model is
evaluated in Appendix~\ref{app:hille-hardy} using the Hille--Hardy formula
for Laguerre modes.}
and contains
$e^{Q_\tau}$, with $Q_\tau$ given by \eqref{eq:gaussian-form}.
The positive exponent
results from dividing out the kinematic prefactor \eqref{eq:prefactor}. By \eqref{eq:slope},
its analogue for $\normalB$, and the last form of \eqref{eq:gaussian-form}, restoring it gives
$\cP U_{+,1/2}=\pi\bigl[k_{t_{31}}(x_{31})k_{t_{42}}(x_{42})+k_{t_{41}}(x_{41})k_{t_{32}}(x_{32})\bigr]$
at all positions, with the heat kernel $k_\ell$ defined in \eqref{eq:bridge-density}: the two terms of
\eqref{eq:free-full} are the two Wick pairings.

At the same eigenvalue $\Lambda=-3$, the even $q=1$ branch has the collinear value
\begin{equation}\label{eq:h-one}
 B_{+,1}(\tau)=\tau\hyp{2}{1}(1,1;3/2;\tau)
 =\sqrt{\frac\tau{1-\tau}}\arcsin\sqrt\tau,
\end{equation}
whereas the canonical free-fermion Wick block is the odd branch
(see also \eqref{eq:wick})
\begin{equation}\label{eq:free-fermion}
 B_{-,1}(\tau)=2\sqrt\tau\left[(1-\tau)^{-1/2}-1\right],\qquad
 U_{-,1}=2\sqrt\tau\left[\frac{e^{Q_\tau}}{\sqrt{1-\tau}}
 -e^{\normalA^2+\normalB^2}\right].
\end{equation}
The full spatial dependence of the even $q=1$ block is given in
\eqref{eq:B-unlike-closed}.
These are three different OPE choices at $\Lambda=-3$. For numerical use, the coefficients $b_1^{(+)}$ and $y_0^{(+)}$
have the regular expressions $b_1^{(+)}=2q^3/(4q-1)$
and $y_0^{(+)}=4q^2$.

Finally, for the even branch at generic $\Delta=2q>1$,
\begin{equation}\label{eq:collision}
 B_{+,\Delta/2}(\tau)\sim K_\Delta(1-\tau)^{-1/2},\qquad
 K_\Delta=\frac{\pi\Gamma(\Delta-\tfrac12)\Gamma(\Delta-1)}
 {\Gamma(\Delta/2)^3\Gamma((\Delta-1)/2)}.
\end{equation}
This follows directly by summing the leading large-$n$ behavior
$b_n^{(+)}\sim(K_\Delta/\sqrt\pi)n^{-1/2}$
obtained from \eqref{eq:coefficients}.
At $\Delta=1$, we take the limit of the ratio of the divergent
Gamma factors, obtaining $K_1=1/2$, while $K_2=\pi/2$ at $\Delta=2$.
The amplitude $K_\Delta$ of the leading collinear singularity at $\tau=1$
is a first crossed-channel datum; the local expansion in this channel may be useful
for deriving bootstrap constraints, as discussed in Section~\ref{sec:connection-moments}.

For the odd branch, the same large-$n$ argument applied to
\eqref{eq:odd-coefficients} gives
\begin{equation}\label{eq:odd-collision}
 B_{-,q}(\tau)\sim K_{-,q}(1-\tau)^{-1/2},\qquad
 K_{-,q}=\frac{\pi\,\Gamma(\Delta-\tfrac12)\,\Gamma(\Delta)}
 {2\,\Gamma(\tfrac{\Delta+1}2)^3\,\Gamma(\tfrac\Delta2)},\quad \Delta=2q.
\end{equation}
We write $K_{+,q}=K_{2q}$ for the even coefficient.

To fix the overall block normalization, we use the identity contribution
in the neutral OPE of $\Psi(t_3,x_3)$ and $\overline\Psi(t_2,x_2)$.\footnote{Beyond the identity contribution, a neutral OPE can involve
``alien'' operators that are neither Schr\"odinger primaries nor descendants
\cite{BekaertMeunierMoroz,GolkarSon}. It would be interesting to develop a full
crossed-channel decomposition that incorporates their pyramid representations
\cite{Boisvert2026}.}
With the external two-point functions
normalized as the unit-mass heat kernels $k_\ell$ defined in \eqref{eq:bridge-density},
this contribution is $k_{t_{32}}(x_{32})k_{t_{41}}(x_{41})$.
On the collinear representative $x_i=0$, the ratio of this contribution to $\cP$ is
$\sqrt{\tau}/(2\pi\sqrt{1-\tau})$.
For branches with finite, nonzero $K_{\chi,q}$, we choose the normalization
so that each block $W_{\chi,q}$ has this leading collinear singularity
as $t_3\to t_2^+$, or equivalently
\begin{equation}\label{eq:normalized-collision}
 \lim_{\tau\to1^-}\sqrt{1-\tau}\,\cN_{\chi,q}B_{\chi,q}(\tau)
 =\cN_{\chi,q}K_{\chi,q}=\frac1{2\pi}.
\end{equation}
This condition fixes $\cN_{\chi,q}=1/(2\pi K_{\chi,q})$ independently of
a dynamical realization. It normalizes the basis blocks; importantly, the weights $c_i$ in
the full correlator \eqref{eq:prefactor} remain theory-dependent.

For a finite sum in \eqref{eq:prefactor}, this convention gives
$\lim_{\tau\to1^-}\sqrt{1-\tau}\,(W/\cP)|_{\normalA=\normalB=0}
=(2\pi)^{-1}\sum_{\mathcal O_i} c_i$.
If the identity contribution gives the leading term in this limit, matching its
coefficient requires
\begin{equation}\label{eq:identity-weight-sum}
 \sum_{\mathcal O_i} c_i=1.
\end{equation}
This identity-matching sum rule is satisfied by the examples of
Sections~\ref{sec:calogero-model} and~\ref{sec:dla}.
Their charged-channel decompositions contain one and two primary
families, respectively, with the infinite towers of relative-motion
descendants summed within the corresponding blocks
as shown in Appendix~\ref{app:hille-hardy}.

\section{\texorpdfstring{Physical realizations and a simple bootstrap argument}{Physical realizations and a simple bootstrap argument}}\label{sec:realizations}
We now discuss a few
simple dynamical realizations of the blocks. A general charged four-point function can
contain blocks with different Casimir eigenvalues. In each example below the
contributing blocks share one eigenvalue, but more than one block may contribute.

In the second-quantized Calogero model the blocks reproduce the known four-point
function and give its fermionic counterpart (Sections~\ref{sec:calogero-model}
and~\ref{sec:free-limits}). In the diffusion-limited reaction $A+B\to\emptyset$, a single
correlator contains both even blocks of a shadow pair at the same eigenvalue
(Section~\ref{sec:dla}). 
In Section~\ref{sec:dla-signed} we use
the nonnegativity of the unlike-species response to fix a mixing coefficient that its
collinear restriction leaves free.

\subsection{The Calogero model and the Casimir equation}\label{sec:calogero-model}
For the second-quantized Calogero model we identify \cite[Eqs.~(2.2)--(2.3), (4.5)]{Shimada}
\begin{equation}\label{eq:calogero-identification}
 \Delta=\nu+\tfrac32,\qquad h=\frac\nu2+\frac34,\qquad
 g=\nu^2-\tfrac14,\qquad \Lambda_\Delta=4g-3.
\end{equation}
The symmetric, or bosonic, extension corresponds to $U_{+,h}$. With unit-mass elementary kernels its
four-point function is
\begin{equation}\label{eq:normalization}
 W_+=\cN_{+,h}\cP U_{+,h},\qquad
 \cN_{+,h}=\frac{\Gamma(h)^2}{2^{2h-1}\pi^{3/2}\Gamma(2h-\tfrac12)}.
\end{equation}
With $\cF_h=\tau^{-h}U_{+,h}$, \eqref{eq:normalization}
agrees with the bosonic four-point function of
Ref.~\cite[Eq.~(3.23)]{Shimada}.
The collinear function $\tau^{-h}B_{+,h}$ agrees with
\cite[Eq.~(3.28)]{Shimada}.

The antisymmetric, or fermionic, extension corresponds to $U_{-,h}$, with the normalization
\begin{equation}\label{eq:odd-normalization}
 W_-=\cN_{-,h}\cP U_{-,h},\qquad
 \cN_{-,h}=\frac{\Gamma(h+\tfrac12)^2}
 {2^{2h-1}\pi^{3/2}\Gamma(2h-\tfrac12)}.
\end{equation}
By the duplication formula, both constants agree with the crossed-channel
normalization \eqref{eq:normalized-collision}.
Each Calogero correlator thus realizes a single block with weight $c_1=1$,
satisfying the condition \eqref{eq:identity-weight-sum} for matching the
neutral-channel identity contribution.
Equations~\eqref{eq:odd-full} and \eqref{eq:odd-collinear} give the fermionic
block and its collinear value. Appendix~\ref{app:gaussian-moments} checks both the function and its normalization
by evaluating the double convolution with the signed two-body kernel.

For the pairwise equal-time configuration $t_1=t_2=0$, $t_3=t_4={\mathcal T}>0$,
put $r=x_{21}$, $s=x_{43}$, and $Z=rs/(2{\mathcal T})$.
For $r,s>0$, $Z$ is the argument of 
the modified Bessel function $I_\nu$
\cite[Sec.~4.1]{Shimada}.
Approaching this configuration from the ordered chart
with $t_{21}=\epsilon\to0^+$, $t_{43}=\epsilon'\to0^+$, and fixed positions and $t_{32}={\mathcal T}$ gives,
for $r,s\ne0$,
\begin{equation}\label{eq:equal-time-scaling}
 \tau\sim\frac{\epsilon\epsilon'}{{\mathcal T}^2},\qquad
 \normalA\sim-\frac{r}{\sqrt{2\epsilon}},\qquad
 \normalB\sim\frac{s}{\sqrt{2\epsilon'}},\qquad z\longrightarrow-Z.
\end{equation}
This differs from $\tau\to0$ at fixed $(\normalA,\normalB)$, where $z\to0$.
The equal-time limit must be taken in the full correlator, including the kinematic prefactor.
The small- and large-$Z$ regimes, which yield the convergent S-channel
and asymptotic T-channel OPE expansions, respectively,
are considered within the equal-time configuration, with exponentially small
U-channel corrections to the latter suggestive of resurgence \cite{Shimada}.

On the pairwise equal-time configuration, the Casimir eigenvalue equation
reduces to a two-body evolution equation. In each pair sum
$X_{12}=X_1+X_2$, the coefficients of $\partial_{t_1}$ and $\partial_{t_2}$
agree at $t_1=t_2$, so the eigenvalue equation \eqref{eq:eigenvalue}
restricts to $t_1=t_2$.
We denote the pairwise equal-time correlator by $W^{\rm eq}_{\chi,h}$
and write $t_{\rm i}$ and $R_{\rm i}=(x_1+x_2)/2$ for the initial pair's
time and center. On this configuration, \eqref{eq:omega} with $M_{12}=-2$
gives $\widetilde C_4^{(12)}=16\Omega_{12}$, where
$\Omega_{12}=r^2(\partial_{t_{\rm i}}+\tfrac14\partial_{R_{\rm i}}^2
+\partial_r^2)-\tfrac34$.
By translation invariance, $W^{\rm eq}_{\chi,h}$ depends on $t_{\rm i}$
and $R_{\rm i}$ only through $\mathcal T=t_3-t_{\rm i}$ and $R$.
With $\partial_{t_{\rm i}}=-\partial_{\mathcal T}$ and
$\partial_{R_{\rm i}}=-\partial_R$, the eigenvalue equation gives,
away from $r=0$,
\begin{equation}\label{eq:equal-time-evolution}
 \left[\partial_{\mathcal T}-\frac14\partial_R^2-\partial_r^2+
 \frac{g}{r^2}\right]W_{\chi,h}^{\rm eq}=0,\qquad
 R=\frac{x_3+x_4-x_1-x_2}{2},\qquad g=\frac{\Lambda_\Delta+3}{4}.
\end{equation}
This is the imaginary-time Schr\"odinger equation for two particles
with the inverse-square interaction $g/r^2$.
The Casimir eigenvalue fixes $g$, which is the same for the shadow
dimensions $\Delta$ and $3-\Delta$, related by $\nu\mapsto-\nu$.
\footnote{By \eqref{eq:omega}, $\Omega_{12}$ is the quadratic Casimir
of the intrinsic $\mathfrak{sl}_2$ in the normalization of
Section~\ref{sec:generators}.
In conformal quantum mechanics, the inverse-square coupling $g$
gives the quadratic Casimir value $g/4-3/16$ \cite{DFF},
consistent with the eigenvalue $\Lambda_\Delta/16$ of
$\Omega_{12}/4$, the intrinsic Casimir in that normalization.}
For $\Delta\ne3/2$, the contact condition at $r=0$ distinguishes these
branches: the branch of dimension $\Delta$ behaves as $r^{\Delta-1}$ as
$r\to0^+$.\footnote{At $\Delta=3/2$, the powers coincide, and the independent 
local behaviors are $r^{1/2}$ and $r^{1/2}\log r$
(cf. Section~\ref{sec:shadow}).}
The extension to $r<0$ is fixed by the primary parity $\chi$, even for bosons
and odd for fermions (see the table in Section~\ref{sec:free-limits} for $g=0$).
The Casimir equation thus governs the two-particle evolution in the
middle interval. The endpoint equations govern the free propagation
in the two outer intervals and reconstruct the general-time correlator
from $W^{\rm eq}_{\chi,h}$ through the double convolution
\eqref{eq:B-convolution}.\footnote{For $\beta=2\nu+1\ge1$, the two-body Calogero kernel on an
ordering sector is related by a Vandermonde similarity transformation
to the transition density of Dyson Brownian motion~\cite{Dyson1962}.
This kernel can also be expressed in terms of the $\beta$-generalized
Harish-Chandra--Itzykson--Zuber integral of random matrix theory
\cite{BrezinHikami2003,KatoriTanemura2003}.}

The relative Hamiltonian in \eqref{eq:equal-time-evolution} is
$H_{\nu,0}=-\partial_r^2+(\nu^2-\tfrac14)/r^2$.
To connect this two-body problem with the state--operator map \cite{NishidaSon},
we introduce a harmonic trap of frequency $\omega>0$.
The eigenstates are expressed in terms of Laguerre
polynomials and have energies $E_n=\omega(2n+\nu+1)$
with $n=0,1,\ldots$, where
the Hille--Hardy formula
\cite[Eq.~(18.18.27)]{DLMF} can be used to evaluate the spectral sum for the
relative-motion propagator (Appendix~\ref{app:hille-hardy}).

Since each Calogero correlator is a single charged block, this gives a
resummation of the relative-motion contribution to its pairwise equal-time
boundary value. The center-of-mass oscillator contributes
$E_j^{\mathrm{CM}}=\omega(j+1/2)$, with $j=0,1,\ldots$.
Adding this to the relative energy $E_n$ gives the total energy
$E_{j,n}=E_n+E_j^{\mathrm{CM}}=\omega(\Delta+j+2n)$.
Including the center-of-mass zero-point energy thus gives
$E_{0,0}/\omega=\nu+3/2$, reproducing the primary dimension $\Delta$
in \eqref{eq:calogero-identification} through the state--operator map.
The excited states correspond to its descendants, with dimensions
$E_{j,n}/\omega=\Delta+j+2n$.

\subsection{\texorpdfstring{Four contact branches at $g=0$}{Four contact branches at g=0}}\label{sec:free-limits}
At $g=0$ the relative motion is free away from contact, and the contact condition
distinguishes four scale-invariant realizations.
These realize the four branches of Section~\ref{sec:shadow} at $h=1/2$
and $\widetilde h=1$, in the same order as in its table.
These contact conditions enter the four-point function through the intermediate
two-particle propagator. Let $r$ denote the signed separation of the two particles at
equal time. The table lists the leading behavior of their relative wave function
as $r\to0$, up to an overall factor.
The braces mark the shadow pairs under $h\mapsto3/2-h$ at fixed primary
parity, as in \eqref{eq:shadow}.
\begin{center}
 \small
 \setlength{\tabcolsep}{4pt}
 \usetikzlibrary{calc,decorations.pathreplacing}
 \newcommand{\shadowref}[2]{%
  \tikz[remember picture,baseline=(#1.base)]
   \node[inner sep=0pt,outer sep=0pt] (#1) {\eqref{#2}};}
 \begin{tabular}{@{}lllccl@{\hspace{10pt}}}\toprule
 Realization & $\nu$ & Primary branch & Contact form & Full block & Collinear block\\\midrule
 Free boson & $-1/2$ & Even, $h=1/2$ & $1$ & $U_{+,1/2}$ & \shadowref{shadowone}{eq:free-collinear}\\
 Hard-core boson & $+1/2$ & Even, $h=1$ & $\lvert r\rvert$ & $U_{+,1}$ & \shadowref{shadowtwo}{eq:h-one}\\
 \addlinespace[5pt]
 Fermionic Tonks--Girardeau & $-1/2$ & Odd, $h=1/2$ & $\operatorname{sgn}(r)$ & $U_{-,1/2}$ & \shadowref{shadowthree}{eq:h-one}\\
 Free fermion & $+1/2$ & Odd, $h=1$ & $r$ & $U_{-,1}$ & \shadowref{shadowfour}{eq:free-fermion}\\\bottomrule
 \end{tabular}
 \begin{tikzpicture}[remember picture,overlay]
  \draw[decorate,decoration={brace,amplitude=3pt}]
   ($(shadowone.north east)+(4pt,1pt)$) --
   ($(shadowtwo.south east)+(4pt,-1pt)$);
  \draw[decorate,decoration={brace,amplitude=3pt}]
   ($(shadowthree.north east)+(4pt,1pt)$) --
   ($(shadowfour.south east)+(4pt,-1pt)$);
  \node[anchor=west,align=center,inner sep=0pt]
   at ($(shadowone.north east)!0.5!(shadowfour.south east)+(13pt,0)$)
   {Shadow\\pairs};
 \end{tikzpicture}
\end{center}
The fermionic Tonks--Girardeau (FTG) branch is the limit of an attractive
zero-range odd-wave interaction, whose relative wave function approaches opposite
nonzero limits on the two sides of contact \cite{GirardeauMinguzzi}.
It is Bose--Fermi dual to the ideal Bose gas, just as the hard-core Bose gas is dual
to the free Fermi gas.

With the normalizations $\cN_{+,1/2}=1/\pi$ and
$\cN_{-,1}=1/(4\pi)$, the free-boson and free-fermion
four-point functions take the Wick form
\begin{equation}\label{eq:wick}
 \begin{aligned}
 W_{\mathrm{free},\mathrm{B/F}}
 =k_{t_{41}}(x_{41})k_{t_{32}}(x_{32})
 &{}\mathbin{\pm}k_{t_{31}}(x_{31})k_{t_{42}}(x_{42}),\\[4pt]
 \left.W_{\mathrm{free},\mathrm{B/F}}\right|_{x_i=0}
 =\frac1{2\pi}\biggl[\frac1{\sqrt{t_{41}t_{32}}}
 &{}\mathbin{\pm}\frac1{\sqrt{t_{31}t_{42}}}\biggr].
 \end{aligned}
\end{equation}

By \eqref{eq:resonance-difference},     
the hard-core and FTG branches share the same
collinear block, the arcsine \eqref{eq:h-one}, but have different full blocks. The free
fermion has yet another collinear block \eqref{eq:free-fermion}. All four realizations have
the same Casimir eigenvalue $\Lambda=-3$.

\subsection{\texorpdfstring{The diffusion-limited reaction $A+B\to\emptyset$}{The diffusion-limited reaction A+B -> 0}}\label{sec:dla}
Consider the diffusion-limited reaction $A+B\to\emptyset$ on a line.
Both species diffuse with diffusion constant $1/2$.
Since the reaction is a relevant perturbation in one dimension, any
nonzero finite microscopic reaction rate leads to the diffusion-limited fixed point
in the long-distance, long-time limit \cite{LeeCardy}.
We work at this fixed point, where particles of unlike species annihilate with
certainty upon first encounter, while particles of the same species do not react.

We study an ordered vacuum response. Starting from the empty state, inject one
particle at $(t_1,x_1)$ and another at $(t_2,x_2)$; remove a particle at $(t_3,x_3)$ and
detect the remaining particle at $(t_4,x_4)$, with $t_1<t_2<t_3<t_4$.
In the Doi--Peliti formalism \cite{Doi,Peliti}, the sum over species at each insertion
is described by the normalized total fields $\Phi=(a+b)/\sqrt2$ and
$\widetilde\Phi=(\tilde a+\tilde b)/\sqrt2$.\footnote{Here $a,b$ are annihilation
fields and $\tilde a=a^*-1$, $\tilde b=b^*-1$ are shifted creation, or response, fields
\cite[Eq.~(2.17)]{LeeCardy}.}
With this normalization, $\langle\Phi\widetilde\Phi\rangle$ is the heat kernel
$k_\ell$ defined in \eqref{eq:bridge-density}. The corresponding four-point function is
$
W_{A+B}=\langle\Phi(t_4,x_4)\Phi(t_3,x_3)
\widetilde\Phi(t_2,x_2)\widetilde\Phi(t_1,x_1)\rangle_{\mathrm{vac}}.
$

Only trajectories in which both injected particles survive until $t_3$ contribute.
As the reaction cannot create particles, if they annihilate earlier, the system
remains empty and the later insertions give zero.
As a result, along a contributing trajectory, there is one particle between $t_1$ and $t_2$, two between $t_2$ and $t_3$, and one
between $t_3$ and $t_4$.

Expanding the product of the four total fields gives a weight of $1/2$ to each of the
same-species and unlike-species contributions. The same-species
contribution is the \emph{free-boson four-point function}.
For the unlike-species contribution, only trajectories
in which the two particles do not meet during the middle interval contribute,
which amounts to giving the Brownian motion of their signed separation $r$
an absorbing boundary at $r=0$.\footnote{The survival problem for arbitrary numbers of species and walkers,
with annihilation only between different species, was studied in
\cite{CardyKatori}.}

By the Karlin--McGregor formula \cite{KarlinMcGregor}, the transition density
for the unlike-species pair to survive without meeting is a determinant of free heat
kernels in each ordering sector. When we sum the contributions with either species
removed at $t_3$ and the other detected at $t_4$, the two assignments of the final
positions to the species give the symmetric hard-core boson kernel of Girardeau's
mapping \cite[Eqs.~(3)--(4)]{Girardeau}. Appendix~\ref{app:unlike} gives the density,
its ordering conditions, and this symmetrization.
Together with the free one-particle propagation in the outer intervals, this gives
the \emph{hard-core boson four-point function}. The two channels are then
the two even realizations in the table of Section~\ref{sec:free-limits}, and the full four-point function is
\begin{equation}\label{eq:dla}
 W_{A+B}=\tfrac12\,\cP\bigl[\cN_{+,1/2}U_{+,1/2}(\tau,\normalA,\normalB)
 +\cN_{+,1}U_{+,1}(\tau,\normalA,\normalB)\bigr],
\end{equation}
for arbitrary real spatial positions at the stated ordered times. Here $\cP$ is the
kinematic prefactor defined in \eqref{eq:prefactor}, and the normalization constants in
\eqref{eq:normalization} give $\cN_{+,1/2}=1/\pi$ and $\cN_{+,1}=1/\pi^2$.
In the decomposition \eqref{eq:prefactor}, the free (same-species) and hard-core
(unlike-species) blocks have weights $c_1=c_2=1/2$, fixed by the normalized total
fields above. The sum $c_1+c_2=1$ satisfies the condition
\eqref{eq:identity-weight-sum} for matching the neutral-channel identity contribution.
The free-boson block is obtained from \eqref{eq:free-full} as
$U_{+,1/2}=\sqrt\tau\,\cF_{+,1/2}$.
Appendix~\ref{app:unlike} gives the closed form \eqref{eq:B-unlike-closed}
for the hard-core boson block $U_{+,1}$. The two blocks have dimensions $\Delta=1$ and $\Delta=2$ and the common
Casimir eigenvalue $\Lambda=-3$.

The one-particle propagation in the outer intervals is free, and hence the endpoint equations
\eqref{eq:free-endpoints} hold. At the diffusion-limited fixed point the middle-interval
interaction is the absorbing condition at coincidence. 
Since this condition is preserved by
Galilean transformations, dilatations, and special conformal transformations, 
the four-point response is Schr\"odinger covariant.\footnote{The half-line problem with a vanishing
boundary condition treated using Schr\"odinger symmetry in
\cite[Eqs.~(3.35)--(3.38)]{Henkel}, 
where a fixed wall breaks Galilean invariance,
corresponds to the relative motion of the
surviving unlike-species pair in each ordering sector ($r>0$ or $r<0$). 
The fixed wall in that problem is represented here by the contact boundary in
the relative coordinate.}

Dividing \eqref{eq:dla} by the kinematic prefactor and then restricting
to the collinear locus gives
\begin{equation}\label{eq:dla-collinear}
\left.\cP^{-1}W_{A+B}\right|_{\normalA=\normalB=0}
 =\frac1{2\pi}\Bigl[B_{+,1/2}(\tau)+\frac1\pi B_{+,1}(\tau)\Bigr].
\end{equation}
For the representative $x_i=0$ for all $i$,
$\cP^{-1}=\sqrt{t_{21}t_{43}}$.
On the collinear locus, the hard-core contribution $\cN_{+,1}\cP B_{+,1}$
is the crossed free contraction $k_{t_{32}}(x_{32})k_{t_{41}}(x_{41})$ multiplied by
$(2/\pi)\arcsin\sqrt\tau$. This factor is the correlation between the signs of a
Brownian bridge at two times \cite[Eq.~(5.34)]{Godreche}, and Appendix~\ref{app:unlike}
obtains it as $G_\tau(0,0)$.

\subsection{Nonnegativity beyond collinearity selects the even shadow branch}\label{sec:dla-signed}
For collinear insertions, the convolution of the Karlin--McGregor kernel
\cite{KarlinMcGregor} with the outer heat kernels reduces to a Brownian-bridge sign
correlation and gives the normalized unlike-species response $B_{+,1}$
(Appendix~\ref{app:unlike}). We use this collinear result and the Casimir value
$\Lambda=-3$ for free relative motion away from contact
(Section~\ref{sec:free-limits}) as input to a simple bootstrap argument, in which
nonnegativity of the unlike-species response fixes its full block.

At this eigenvalue, the even block with $\Delta=2$ and the odd block with
$\Delta=1$ share the same collinear value. 
Among solutions of the two endpoint equations and of the Casimir equation
at this eigenvalue that are analytic in $(\normalA,\normalB)$ and have
the simultaneous spatial inversion symmetry \eqref{eq:U-even}, 
the only freedom at fixed $B=B_{+,1}$ is to add a constant multiple
of the invisible solution of
Section~\ref{sec:resonance}. Using \eqref{eq:resonance-difference}, we write the
general candidate for the unlike-species block as
\begin{equation}\label{eq:dla-mixture}
 U_\kappa=(1-\kappa)U_{+,1}+\kappa U_{-,1/2}
 =U_{+,1}+4\kappa U_{\mathrm{inv}},\qquad \kappa\in\mathbb R.
\end{equation}
By \eqref{eq:resonance-difference}, the seeds of $U_\kappa$ are $B_{+,1}$ and
$Y_{+,1}+4\kappa$. The collinear value is independent of $\kappa$, which enters only
through the mixed seed.

The unlike-species response is an integral over surviving diffusion paths
with nonnegative weights. Since the kinematic prefactor is positive, its reduced block must satisfy
$U_\kappa(\tau,\normalA,\normalB)\ge0$ for every $0<\tau<1$ and every
$(\normalA,\normalB)\in\mathbb R^2$, that is, throughout the ordered real
shape space $\mathcal M_4^\uparrow$.

Consider the limit $\tau\to0$ at fixed $(\normalA,\normalB)$, in which the lowest
exchanged dimension dominates. The invisible solution is exactly
$U_{\mathrm{inv}}=\sqrt\tau\,O_{1/2}(\normalA)O_{1/2}(\normalB)$ by
\eqref{eq:invisible-lift} and \eqref{eq:even-odd}, and $4U_{\mathrm{inv}}$ is the leading
term of the odd block with $\Delta=1$ in \eqref{eq:odd-full}. Since the even block
$U_{+,1}$ begins at order $\tau$ by \eqref{eq:full-block}, the second term of
\eqref{eq:dla-mixture} dominates:
\begin{equation}\label{eq:dla-ope-limit}
 \lim_{\tau\to0}\tau^{-1/2}U_\kappa(\tau,\normalA,\normalB)
 =4\kappa\,O_{1/2}(\normalA)\,O_{1/2}(\normalB).
\end{equation}
Since $O_{1/2}(\normalA)=\normalA\,\hyp{1}{1}(1;\tfrac32;\normalA^2)$ has the sign
of $\normalA$, the limit has the sign of $\kappa\normalA\normalB$, that is, of $\kappa z$.
For any $\kappa\ne0$, we choose
$\normalA,\normalB$ so that $\kappa\normalA\normalB<0$.
Then $U_\kappa<0$ for all sufficiently small positive $\tau$,
contradicting nonnegativity throughout $\mathcal M_4^\uparrow$.
It follows that
\begin{equation}\label{eq:dla-nonnegativity}
 \kappa=0,\qquad U_\kappa=U_{+,1}.
\end{equation}
Thus nonnegativity recovers the primary parity lost under the collinear projection.

The physical reason is that absorption at contact makes the pair propagator vanish when
the two particles meet. In a two-body realization this is the behavior of the even
block $U_{+,1}$ with $\Delta=2$ ($\nu=1/2$), whose relative wave function vanishes
linearly at contact.
The odd block $U_{-,1/2}$
with $\Delta=1$ ($\nu=-1/2$ with the odd extension)
corresponds to the FTG limit \cite{GirardeauMinguzzi}:
its relative wave function has opposite nonzero one-sided limits at contact.
This branch has a quantum realization, but \eqref{eq:dla-nonnegativity} rules out any contribution from it to the unlike-species response considered here.
We conclude that the unlike-species response belongs to the even
$\Delta=2$ branch, the shadow of the free same-species branch at $\Delta=1$.
Both branches occur with positive weights in the total response,
giving a physical realization of the shadow pair discussed in
Section~\ref{sec:shadow}.

\Needspace{12\baselineskip}
\section{Discussion and outlook}\label{sec:discussion}
\subsection{Main results}\label{sec:results-scope}
We have constructed
charged Schr\"odinger blocks for ordered stochastic responses and vacuum correlators,
under the assumptions of Section~\ref{sec:ordered-channel}, independently of any dynamical realization.
The full quartic
Casimir problem is equivalent to a closed system for the temporal seeds $(B,Y)$, and the endpoint
equations reconstruct the spatial dependence via the exact Kummer lift. 
At generic exchanged dimension, the OPE conditions fix a unique block
up to normalization for each primary parity at each of the shadow-related
dimensions $\Delta$ and $3-\Delta$.
The bosonic and fermionic Calogero four-point functions provide
concrete realizations of the even- and odd-primary blocks, respectively.
Indeed, at pairwise equal times and away from contact, the quartic
Casimir equation reduces to the two-body Calogero evolution equation
\eqref{eq:equal-time-evolution} with the coupling fixed by the Casimir eigenvalue.
The endpoint equations then reconstruct the general-time correlator
from the equal-time Calogero kernel through the double convolution
\cite{Shimada}, whose evaluation provides an independent check of
the block expressions and normalizations
(Appendix~\ref{app:integral}).

At the one-dimensional diffusion-limited fixed point of $A+B\to\emptyset$,
the four-point response summed over particle species contains both even members
of the shadow pair, with dimensions $\Delta=1,2$ and positive weights.
Nonnegativity of the unlike-species response throughout the ordered real
shape space resolves a primary-parity ambiguity that remains in its complete
collinear function. The mixed seed retains the information lost in the collinear
restriction, and its contribution away from collinearity makes this selection possible.

The weights with which these blocks enter a general correlator remain
theory-dependent. Determining or constraining these weights through consistency
between channels and physical conditions such as nonnegativity is a central goal
of a Schr\"odinger bootstrap, and in particular of its application to four-point
response functions in nonequilibrium systems.

\subsection{Towards a Schr\"odinger bootstrap: local connection data}\label{sec:connection-moments}
The collinear limit $t_3\to t_2^+$, or $\tau\to1$, probes the neutral OPE
of the two middle insertions. We have used its identity contribution to
normalize the charged blocks as in \eqref{eq:normalized-collision}.
The collision coefficients in \eqref{eq:collision} and \eqref{eq:odd-collision}
give the leading terms of the local connection expansion at $\tau=1$, that is,
of the expansion near $\tau=1$ of the collinear blocks defined by their OPE
behavior at $\tau=0$.
For either primary parity the collinear $\hyp{4}{3}$ has parametric excess
$-1/2$, the sum of its lower parameters minus the sum of its upper
parameters, giving the singular power $(1-\tau)^{-1/2}$ at generic parameters.
At generic parameters the N{\o}rlund--B\"uhring expansion
\cite{Norlund,Buhring,BuhringSrivastava,KarpPrilepkina} takes the form
\begin{equation}\label{eq:connection}
B_{\chi,q}(1-w)=K_{\chi,q}\,w^{-1/2}\sum_{n\ge0}\sigma_n(\Lambda_\Delta)\,w^n
 +H_{\chi,q}(w),\qquad \sigma_0=1,\quad w=1-\tau,
\end{equation}
where $\Delta=2q$ and $H_{\chi,q}$ is analytic at $w=0$.
Substituting $B\sim w^\lambda$ into \eqref{eq:fourth-order} gives the indicial equation
$\lambda(\lambda-1)(\lambda-2)(\lambda+\tfrac12)=0$ at $w=0$.
The exponent $-\tfrac12$ is nonresonant with the other three exponents,
since its differences from them are not integers, so its Frobenius series
is unique up to the overall coefficient $K_{\chi,q}$ at fixed Casimir.
Its coefficients $\sigma_n$ depend on the exchanged family only through
$\Lambda_\Delta$, starting with $\sigma_0=1$ and
$\sigma_1(\Lambda_\Delta)=(\Lambda_\Delta+1)/4$.

In the physical examples of Section~\ref{sec:realizations}, each correlator
involves only one Casimir eigenvalue. At fixed coupling, the Calogero correlator
contains a single block, while the reaction response contains two even blocks
with $\Delta=1,2$ and the common eigenvalue $\Lambda=-3$.
In each example, all contributing blocks therefore have the same Casimir
eigenvalue: the positive measure on Casimir eigenvalues defined by the block
weights is concentrated at one point.
The two reaction blocks form a shadow pair with the same Casimir eigenvalue,
so their weights enter every moment of this measure only through their sum.

More general correlators, however, can receive contributions from primary
families with distinct Casimir eigenvalues. 
Natural candidates include multicomponent systems with different
inverse-square couplings in different pair channels, such as
spin Calogero--Sutherland models in their rational limit
\cite{HaHaldane} and reaction--diffusion models with long-range
reaction rates \cite{TauberHowardVollmayrLee,ParkDeem}.
For positive block weights,
the local connection expansion \eqref{eq:connection} may then yield an
\emph{infinite hierarchy of moment inequalities} for the corresponding positive
measure on Casimir eigenvalues.
A related formulation in terms of positive OPE measures has been developed
for relativistic CFTs \cite{PolandRogelberg}.
The constraints obtained from these moment inequalities could be strengthened
further by incorporating independent crossed-channel data.
Obtaining such data calls for further analysis of the neutral sector,
whose representation theory has been studied in
\cite{Boisvert2025,Boisvert2026}. Determining the full neutral-channel
decompositions of these correlators and relating them to the charged-channel
expansion remain open problems.

The full spatial dependence can provide constraints that are invisible
on the collinear locus. 
Section~\ref{sec:dla-signed} already gives a concrete example:
for a specified collinear response and Casimir eigenvalue, nonnegativity
throughout the ordered real shape space fixes the mixing coefficient
left undetermined by the complete collinear function.
This motivates a bootstrap analysis that combines the local connection
data with nonnegativity of the full spatially dependent response.

\subsection{Factorized collinear blocks and higher-dimensional extensions}\label{sec:two-body-extensions}
The even- and odd-primary collinear blocks both admit factorized
representations that may be useful for bootstrap analyses. For example,
Bailey's product formula \cite{Bailey,DLMF} gives the even-primary block as
\begin{equation}\label{eq:bailey-outlook}
 B_{+,q}(\tau)=\frac{\tau^2}{q-1}\frac d{d\tau}
 \Bigl[\tau^{q-1}\hyp{2}{1}(q-1,q;\tfrac12;\zeta)\,
 \hyp{2}{1}(q-1,q;2q-\tfrac12;\zeta)\Bigr],
 \quad \zeta=\frac{1-\sqrt{1-\tau}}2.
\end{equation}
At $q=1$, parameter continuation gives \eqref{eq:h-one}.
The variable $\zeta$ is related to the radial coordinate of
Hogervorst and Rychkov \cite{HogervorstRychkov}, evaluated at the temporal
cross-ratio, by $\zeta=\rho/(1+\rho)$ and $\tau=4\rho/(1+\rho)^2$.
Although a detailed analysis of this factorized representation is beyond
the scope of this paper, the two ${}_2F_1$ factors suggest an interpretation
in terms of center-of-mass and relative motion, up to the displayed derivative.
This motivates the study of collinear partial waves of
inverse-square two-body systems in higher spatial dimensions, where
angular-momentum decomposition leads to radial inverse-square Hamiltonians.
In the embedding of complexified Lie algebras
$(\mathfrak{schr}_d)_{\mathbb C}\subset
(\mathfrak{conf}_{d+2})_{\mathbb C}$ \cite{HenkelUnterberger},
the Schr\"odinger mass corresponds to the momentum conjugate to 
an auxiliary null coordinate.
Projection onto fixed-mass Fourier modes, used in constructions of
nonrelativistic holographic correlation functions \cite{VolovichWen},
may also relate these collinear waves to relativistic conformal partial waves.

For full blocks at general spatial configurations in more than one
dimension, the two signed residuals
become vectors. Writing $a=|\boldsymbol\normalA|^2$,
$b=|\boldsymbol\normalB|^2$, and
$c=\boldsymbol\normalA\!\cdot\!\boldsymbol\normalB$, their Gram matrix
$G=\bigl(\begin{smallmatrix}a&c\\c&b\end{smallmatrix}\bigr)$ is positive
semidefinite, so $c^2\le ab$. For $d=1$ its determinant vanishes, $c^2=ab$, reducing
the number of independent continuous invariants among $(\tau,a,b,c)$ from four
to three.%
\footnote{Under the standard Brownian-bridge measure, $G$ has the
Wishart distribution $W_2(d,\Sigma_\tau)$ \cite{Uhlig1994}. For $d=1$, it is
singular at every configuration, whereas its mean $\Sigma_\tau$ of
\eqref{eq:covariance} is not. It would be interesting to see whether, for $d>1$, the temporal
seeds are organized by $\mathrm{SO}(d)$ harmonics in the angle between the velocity kinks
$\boldsymbol\normalA$ and $\boldsymbol\normalB$.}
Whether the endpoint equations again reduce the Casimir problem to
finitely many temporal seeds, and how many seeds are required, remain open
questions.

\section*{Acknowledgements}
The author thanks Tomohiro Sasamoto and Makoto Katori for insightful suggestions, 
Jumpei Hayashi and  
Hidehiko Shimada for discussions and their respective collaborations on related work, and
Malte Henkel for correspondence on their results. 
This work was supported in part by JSPS KAKENHI Grant Number JP24K06892. 
Assistance with checking calculations was provided by Claude Opus 5.5; the author takes full responsibility for the content.

\appendix
\section{Explicit computation of the Casimir identity}\label{app:casimir}
This appendix derives \eqref{eq:operator-identity} directly from the pair generators.
For the generators in \eqref{eq:generators}, we use the pair sums $X_{12}=X_1+X_2$.

\subsection{Differentiate before fixing the remaining scale}\label{app:derivative-rules}
In the frame \eqref{eq:frame}, $T=-t_1$ and $\tau=(t_2-t_1)/T$. Introduce $\widehat x_i=x_i/\sqrt T$ for $i=1,2$.
The derivatives, taken with the other original coordinates fixed, become
\begin{equation}\label{eq:A-derivatives}
 \begin{aligned}
 \partial_{t_1}&=-\partial_T-\frac{1-\tau}{T}\partial_\tau
 +\frac{\framex\partial_\framex+\framey\partial_\framey}{2T},\qquad
 \partial_{t_2}=\frac1T\partial_\tau,\\
 \partial_{x_i}&=T^{-1/2}\partial_{\widehat x_i},\qquad i=1,2.
 \end{aligned}
\end{equation}
On the right, $(T,\tau,\framex,\framey)$ are independent variables. We next change to the signed coordinates and
write the prefactor needed for conjugation in the same variables:
\begin{equation}\label{eq:A-change}
 \begin{gathered}
 \normalA =\frac{(1-\tau)\framex-\framey}{\sqrt{2\tau(1-\tau)}},\qquad
 \normalB =\frac{\framey}{\sqrt{2(1-\tau)}},\\
 \log\cP_\infty=-\frac12\log(T\tau)-\frac{\framey^2}{1-\tau}-\frac{(\framey-\framex)^2}{2\tau}.
 \end{gathered}
\end{equation}
At fixed $T$, the chain rule for the coordinate transformation in \eqref{eq:A-change} gives
\begin{equation}\label{eq:A-chain}
 \begin{aligned}
 \partial_\framex&=\sqrt{\frac{1-\tau}{2\tau}}\partial_\normalA ,\qquad
 \partial_\framey=\frac1{\sqrt{2(1-\tau)}}\left(\partial_\normalB-\frac1{\sqrt\tau}\partial_\normalA\right),\\
 \partial_\tau\big|_{\framex,\framey}&=\partial_\tau\big|_{\normalA ,\normalB }
 -\left(\frac{\normalA}{2\tau(1-\tau)}+\frac{\normalB}{\sqrt\tau(1-\tau)}\right)\partial_\normalA
 +\frac{\normalB}{2(1-\tau)}\partial_\normalB.
 \end{aligned}
\end{equation}
\Needspace{12\baselineskip}
Write each pair generator as $X_{12}=V+c$, where $V$ is its vector part, consisting of the first-order derivative terms,
and $c$ is its multiplication term. We transform $V$ using \eqref{eq:A-derivatives} and \eqref{eq:A-chain}, then set
$\overline X=\cP_\infty^{-1}X_{12}\cP_\infty$. This conjugation leaves $V$ unchanged and replaces $c$ by
$c+V(\log\cP_\infty)$, with $\log\cP_\infty$ given in \eqref{eq:A-change}.
Applying this procedure to the pair sums of the generators in \eqref{eq:generators} gives
\begin{equation}\label{eq:A-generator-table}
 \begin{aligned}
 \overline P&=\frac{-\sqrt\tau\partial_\normalA +\partial_\normalB -4\normalB }{\sqrt{2T(1-\tau)}},\qquad
 \overline G=-\sqrt{\frac{T(1-\tau)}2}\partial_\normalB ,\\
 \overline H&=\frac1T\left[-\theta_T+\theta+
 \frac{\normalB(\partial_\normalB-4\normalB)-(\tau\normalA+2\sqrt\tau\normalB)\partial_\normalA}{2(1-\tau)}\right],\\
 \overline D&=2\theta_T,\qquad
 \overline K=-T\bigl[\theta_T+(1-\tau)(\theta+\tfrac12\normalB \partial_\normalB )\bigr].
 \end{aligned}
\end{equation}
For example, in the $(\framex,\framey)$ variables the vector part of $G_{12}$ is $-\sqrt T[\partial_\framex+(1-\tau)\partial_\framey]$. The logarithmic
derivative of the prefactor cancels its multiplication term $-\sqrt T(\framex+\framey)$, and \eqref{eq:A-chain} then gives $\overline G$.
The same chain rules applied to $-2\partial_{t_1}-\partial_{x_1}^2$ give \eqref{eq:chain-endpoint}. The scale derivative $\theta_T$ must
be retained until all generator products have been formed.

\subsection{Separating the scale dependence}
To write the subtracted generators $H_0$, $D_0$, and $K_0$ of
\eqref{eq:subtractions}, introduce
\begin{equation}\label{eq:A-N}
 N_\normalA =\partial_\normalA ^2-4\normalA \partial_\normalA ,\qquad N_\normalB =\partial_\normalB ^2-4\normalB \partial_\normalB .
\end{equation}
As $M_{12}=-2$, the subtractions in \eqref{eq:subtractions} are additions of one quarter of the indicated quadratic
products. Equation~\eqref{eq:A-generator-table} gives
\begin{equation}\label{eq:A-subtracted}
 \overline H_0=T^{-1}(-\theta_T+\sH),\qquad \overline D_0=2\theta_T+\sD,\qquad
 \overline K_0=T(-\theta_T+\sK),
\end{equation}
where
\begin{equation}\label{eq:A-three}
 \begin{aligned}
 \sH&=\theta+\frac{\tau N_\normalA +N_\normalB -2\sqrt\tau\partial_\normalA \partial_\normalB -4}{8(1-\tau)},\\
 \sD&=\tfrac12+\frac{\sqrt\tau\partial_\normalA \partial_\normalB -N_\normalB }{4},\qquad
 \sK=-(1-\tau)\left(\theta-\frac{N_\normalB }{8}\right).
 \end{aligned}
\end{equation}
These operators are independent of $T$. In the products below, $\theta_T$ acts on the $T^{-1}$ or $T$ in the right-hand factor.
Using $\theta_T T^{-1}=T^{-1}(\theta_T-1)$ and $\theta_T T=T(\theta_T+1)$ gives
\begin{equation}\label{eq:A-products}
 \begin{aligned}
 \overline K_0\overline H_0&=(-\theta_T+1+\sK)(-\theta_T+\sH),\\
 \overline H_0\overline K_0&=(-\theta_T-1+\sH)(-\theta_T+\sK).
 \end{aligned}
\end{equation}
The $\theta_T^2$ terms cancel in the Casimir combination:
\begin{equation}\label{eq:A-omega}
 \begin{aligned}
 \overline\Omega&=\overline D_0^{\,2}-2\{\overline K_0,\overline H_0\}\\
 &=4\theta_T(\sH+\sD+\sK)+\sD^2-2\{\sK,\sH\}+2(\sK-\sH).
 \end{aligned}
\end{equation}
The shifts in \eqref{eq:A-products} leave the term $2(\sK-\sH)$. Since $\theta_T U=0$ and $\theta_T$ commutes with
$\sH,\sD,\sK$, only the term proportional to $\theta_T$ drops out
when $\overline\Omega$ acts on $U$.

\subsection{\texorpdfstring{Expanding the Casimir and isolating the endpoint terms}{Expanding the Casimir and isolating the endpoint terms}}
On functions independent of $T$, substituting \eqref{eq:A-three} into \eqref{eq:A-omega} and applying the product rule
gives the following expression. We use $\theta f(\tau)=f(\tau)\theta+\tau f'(\tau)$ and
$\partial_\normalA\normalA=\normalA\partial_\normalA+1$, with the analogous rule for $\normalB$:
\begin{equation}\label{eq:A-expanded}
 \begin{aligned}
 \overline\Omega={}&4(1-\tau)\theta^2-6\theta+\frac54
 +\frac\tau2(N_\normalA+N_\normalB)\theta\\
 &+\left[\frac\tau4\bigl(\normalA\partial_\normalB+\normalB\partial_\normalA-4\normalA\normalB\bigr)
 -\sqrt\tau\left(\theta-\frac12\right)\right]\partial_\normalA\partial_\normalB.
 \end{aligned}
\end{equation}
The fourth-order derivatives cancel between the individual products, before either endpoint equation is used.
To separate the terms proportional to the endpoint operators, we expand
\begin{equation}\label{eq:A-remove}
 \begin{aligned}
 \frac\tau4\bigl[(2\theta+\normalB \partial_\normalB )L_\normalA +(2\theta+\normalA \partial_\normalA )L_\normalB \bigr]
 &={}\frac\tau2(N_\normalA+N_\normalB)\theta-4\tau\theta^2\\
 &\quad+\frac\tau4\bigl(\normalA\partial_\normalB+\normalB\partial_\normalA-4\normalA\normalB\bigr)
 \partial_\normalA\partial_\normalB.
 \end{aligned}
\end{equation}
All operators act on factors to their right; the leftmost $\tau$ factors are not differentiated.
Subtracting \eqref{eq:A-remove} from \eqref{eq:A-expanded} leaves
\begin{equation}\label{eq:A-remainder}
 4\theta(\theta-\tfrac32)+\frac54-\sqrt\tau(\theta-\tfrac12)\partial_\normalA \partial_\normalB.
\end{equation}
Since $\widetilde C_4^{(12)}=16\Omega$, subtracting the Casimir eigenvalue $4\Lambda_\Delta$ amounts to subtracting
$\Lambda_\Delta/4$ from $\overline\Omega$. The temporal part then factorizes as
\begin{equation}\label{eq:A-final}
 4\theta(\theta-\tfrac32)+\frac{5-\Lambda_\Delta}{4}=4(\theta-h)(\theta+h-\tfrac32).
\end{equation}
Together with \eqref{eq:A-remove}, this proves \eqref{eq:operator-identity}.

\section{Independent check from the Calogero integral}\label{app:integral}
In this appendix, we evaluate the double convolution \cite{Shimada} %
to obtain the triple-series representation of the general four-point function. 
This representation was already presented in \cite[Eqs.~(3.23)--(3.25)]{Shimada}, 
where its detailed derivation was reserved for a separate publication. 
We separate the signs of the relative coordinates and evaluate the resulting Laplace--Mellin transforms 
term by term using the Gamma integral \eqref{eq:B-gamma}, 
making the dependence on the signed invariant $z$ explicit. 
We also obtain the sums of Kummer products directly by evaluating Gaussian moments. 
These evaluations check the series representations and their normalizations 
without using the Casimir equation or endpoint recurrences. 
Finally, we use the Hille--Hardy formula to evaluate the sum over relative-motion descendants 
and explain its relation to the state--operator map. 
Throughout this appendix, we write $\cF_h=\cF_{+,h}$.

\subsection{Split the convolution into two positive-distance integrals}
Let $p=t_{21}$, $q=t_{32}$, and $r=t_{43}$. The single-particle kernel and the bosonic two-particle kernel
are \cite[Eqs.~(3.16), (3.19)--(3.20)]{Shimada}
\begin{equation}\label{eq:B-kernels}
 \begin{aligned}
 K_1(X;Y;t)&=\frac{e^{-(X-Y)^2/(2t)}}{\sqrt{2\pi t}},\\
 K_2(X_1,X_2;Y_1,Y_2;t)&=e^{-[(Y_1-Y_2)^2+(X_1-X_2)^2+(X_1+X_2-Y_1-Y_2)^2]/(4t)}\\
 &\quad\times\sqrt{\frac{|Y_1-Y_2|\,|X_1-X_2|}{4\pi t^3}}\,
 I_\nu\!\left(\frac{|Y_1-Y_2|\,|X_1-X_2|}{2t}\right),
 \end{aligned}
\end{equation}
\begin{equation}\label{eq:B-convolution}
 W=\int_{\mathbb R^2}dx_1'\,dx_4'\,
 K_1(x_4;x_4';r)K_2(x_3,x_4';x_1',x_2;q)K_1(x_1';x_1;p).
\end{equation}
Separate the signs of the two relative positions:
\begin{equation}\label{eq:B-signs}
 x_2-x_1'=\rho_1x,\qquad x_4'-x_3=\rho_4y,\qquad
 x,y>0,\quad \rho_1,\rho_4=\pm1.
\end{equation}
For fixed $y,\rho_4$, put
\begin{equation}\label{eq:B-constants}
 \begin{aligned}
 a^2&=\frac12\left(\frac1p+\frac1q\right),&
 \widetilde a^2&=\frac12\left(\frac1r+\frac1q\right),&c&=\frac{y}{2q},\\
 b&=\frac{x_{21}}p-\frac{x_{32}}q-\rho_4c,&
 \widetilde b&=\frac{x_{43}}r-\frac{x_{32}}q,&a,\widetilde a&>0.
 \end{aligned}
\end{equation}
With these parameters, the Gaussian exponent in \eqref{eq:B-convolution} is
$-a^2x^2+\rho_1bx-S$, where
\begin{equation}\label{eq:B-exponent}
 S=S_0+\widetilde a^2y^2-\rho_4\widetilde b\,y,\qquad
 S_0=\frac{x_{21}^2}{2p}+\frac{x_{32}^2}{q}+\frac{x_{43}^2}{2r}.
\end{equation}
Writing $\mathcal J(b,c)$ for the inner integral at fixed $a$, the convolution splits as
\begin{equation}\label{eq:B-split}
 W=\frac1{(2\pi)^{3/2}q\sqrt{pr}}
 \sum_{\rho_4}\int_0^\infty dy\,e^{-S}\,\mathcal J(b,c),\quad
 \mathcal J(b,c)=\sum_{\rho_1}\int_0^\infty dx\,
 e^{-a^2x^2+\rho_1bx}\sqrt{cx}\,I_\nu(cx).
\end{equation}

\subsection{The first Laplace--Mellin transform}
The first sign sum is $2\cosh(bx)$. Set $u=x^2$ and use
\begin{equation}\label{eq:B-bessel}
 \cosh(bx)=\hyp{0}{1}\!\left(;\tfrac12;\frac{b^2x^2}4\right),\qquad
 I_\nu(cx)=\frac{(cx/2)^\nu}{\Gamma(\nu+1)}
 \hyp{0}{1}\!\left(;\nu+1;\frac{c^2x^2}4\right).
\end{equation}
With $h=\nu/2+3/4$, this gives
\begin{equation}\label{eq:B-laplace}
 \mathcal J(b,c)=\frac{c^{\nu+1/2}}{2^\nu\Gamma(\nu+1)}
 \int_0^\infty du\,u^{h-1}e^{-a^2u}
 \hyp{0}{1}\!\left(;\tfrac12;\frac{b^2u}4\right)
 \hyp{0}{1}\!\left(;\nu+1;\frac{c^2u}4\right).
\end{equation}
The factor two from the sign sum cancels the factor $1/2$ in $dx=du/(2\sqrt u)$.
Expand the two series and apply the elementary Gamma integral
\begin{equation}\label{eq:B-gamma}
 \int_0^\infty du\,u^{s-1}e^{-a^2u}=\frac{\Gamma(s)}{a^{2s}},
 \qquad \operatorname{Re} s>0,
\end{equation}
at $s=h+m+n$. With $\Gamma(h+m+n)=\Gamma(h)(h)_{m+n}$, this Laplace--Mellin transform leads to
\begin{equation}\label{eq:B-humbert}
 \boxed{\mathcal J(b,c)=\frac{c^{\nu+1/2}\Gamma(h)}
 {2^\nu a^{2h}\Gamma(\nu+1)}
 \Psi_2\!\left(h;\tfrac12,\nu+1;\frac{b^2}{4a^2},\frac{c^2}{4a^2}\right)},
\end{equation}
where the Humbert function $\Psi_2$ is defined by the double series
\begin{equation}\label{eq:B-psi2}
 \Psi_2(\alpha;\beta,\gamma;X,Y)=\sum_{m,n\ge0}
 \frac{(\alpha)_{m+n}X^mY^n}{(\beta)_m(\gamma)_n m!n!}.
\end{equation}
Substitution into \eqref{eq:B-split} leaves the second position integral.

\subsection{\texorpdfstring{The second integration: Taylor sectors and the triple series}{The second integration: Taylor sectors and the triple series}}\label{app:second-mellin}
For the remaining integral, write $\rho=\rho_4$ and set $\inty=\widetilde a y>0$
and $c_h=\nu+1=2h-1/2$, taking the positive square root of $\tau$. The relations
\begin{equation}\label{eq:B-positive-scaling}
 \frac1{2qa\widetilde a}=\sqrt\tau,\qquad
 \frac b{2a}=-\normalA-\frac{\rho\sqrt\tau\,\inty}{2},\qquad
 \frac c{2a}=\frac{\sqrt\tau\,\inty}{2},\qquad
 \frac{\widetilde b}{2\widetilde a}=\normalB
\end{equation}
turn \eqref{eq:B-split}--\eqref{eq:B-humbert}, with the normalization
\eqref{eq:normalization}, into
\begin{equation}\label{eq:B-second-integral}
 \cF_h=\frac1{\Gamma(h)}\sum_{\rho=\pm1}\int_0^\infty
 d\inty\,\inty^{2h-1}e^{-\inty^2+2\rho\normalB\inty}\,
 \Psi_2\!\left(h;\tfrac12,c_h;
 (\normalA+\rho\sqrt\tau\,\inty/2)^2,\frac{\tau\inty^2}4\right).
\end{equation}
We first work with $h\ge1/2$, $0<\tau<1$, and finite real $(\normalA,\normalB)$;
absolute convergence is justified by \eqref{eq:B-bound} below.

To separate the two Taylor sectors, introduce the local auxiliary quantities
\begin{equation}\label{eq:B-XZ}
 X=2\normalB\inty,\qquad Z=\frac{2\normalA}{\sqrt\tau\,\inty},\qquad
 (\normalA+\rho\sqrt\tau\,\inty/2)^{2m}
 =\left(\frac{\tau\inty^2}4\right)^m(1+\rho Z)^{2m}.
\end{equation}
The binomial polynomial separates as
\begin{equation}\label{eq:B-parity-polynomials}
 \begin{gathered}
 (1+\rho Z)^{2m}=P_m(Z^2)+2m\rho ZQ_m(Z^2),\\
 P_m(Z^2)=\hyp{2}{1}(-m,\tfrac12-m;\tfrac12;Z^2),\qquad
 Q_m(Z^2)=\hyp{2}{1}(\tfrac12-m,1-m;\tfrac32;Z^2).
 \end{gathered}
\end{equation}
Here $m\ge1$; set $P_0=1$, $Q_0=0$. Since
$e^{\rho X}=\cosh X+\rho\sinh X$, the sign sum keeps the even--even and odd--odd products:
\begin{equation}\label{eq:B-parity-sum}
 \frac12\sum_{\rho=\pm1}e^{\rho X}(1+\rho Z)^{2m}
 =P_m(Z^2)\hyp{0}{1}\!(;\tfrac12;X^2/4)
 +2mXZQ_m(Z^2)\hyp{0}{1}\!(;\tfrac32;X^2/4).
\end{equation}
The arguments and the mixed factor satisfy
\begin{equation}\label{eq:B-signed-invariant}
 \frac{X^2}4=\normalB^2\inty^2,\qquad
 Z^2=\frac{4\normalA^2}{\tau\inty^2},\qquad
 \boxed{XZ=\frac{4z}{\tau}},\qquad
 z=\frac{A_{123}A_{234}}{2t_{31}t_{32}t_{42}}.
\end{equation}
The inverse powers of $\inty$ arising from $(1+\rho Z)^{2m}$ are canceled by the prefactor
$(\tau\inty^2/4)^m$ in \eqref{eq:B-XZ}.
Since $XZ=4z/\tau$, the odd--odd term carries the signed rational invariant
\eqref{eq:z}, independent of the integration distance; its remaining spatial dependence is through
$\normalA^2,\normalB^2$. This gives the $\{1,z\}$--sector decomposition
\eqref{eq:parity-decomposition} even before integration. 
Note, however, that these spatial sectors are distinct 
from primary parity $\chi$: both occur already in the even-primary block.

Expand the finite polynomials and the $\hyp{0}{1}$ functions in \eqref{eq:B-parity-sum}.
In the sector $j=0,1$, let the powers of
$(\normalA,\normalB)$ be $(2\ell+j,2k+j)$ and reindex the $\Psi_2$ series by
$m=M+\ell+j$, $p=M+n$, with $p$ now a summation index.
With $u=\inty^2$, the same Laplace--Mellin transform
\eqref{eq:B-gamma} gives
\begin{equation}\label{eq:B-second-gamma}
 2\int_0^\infty d\inty\,\inty^{2(h+p+k+j)-1}e^{-\inty^2}
 =\Gamma(h+p+k+j).
\end{equation}
After division by $\Gamma(h)^2$, the two Gamma factors give
$h^{2j}(h+j)_{p+\ell}(h+j)_{p+k}$. The triple-series coefficients
\eqref{eq:B-mellin-coefficients} then follow from the finite convolution
\begin{equation}\label{eq:B-mellin-convolution}
 \sum_{M+n=p}\frac1{M!n!(\tfrac12+j)_M(c_h)_n}
 =\frac{4^p(2h-1+j)_{2p}}
 {p!(\tfrac12+j)_p(4h-2+j)_{2p}},\qquad j=0,1.
\end{equation}
The left side is
$\hyp{2}{1}(-p,\tfrac12-j-p;c_h;1)/[p!(\tfrac12+j)_p]$; terminating Vandermonde summation and
\eqref{eq:pochhammer} give the right side.

When points 1, 2, and 3 are collinear, $\normalA=0$ and the two arguments of $\Psi_2$
in \eqref{eq:B-second-integral} coincide. With $u=\inty^2$, resumming the $j=0$ case
of \eqref{eq:B-mellin-convolution} gives the equivalent function identity of
Burchnall and Chaundy \cite[Eq.~(66)]{BurchnallChaundy1941}:
\begin{equation}\label{eq:B-three-collinear}
 \Psi_2\!\left(h;\tfrac12,c_h;\frac{\tau u}4,\frac{\tau u}4\right)
 =\hyp{3}{3}\!\left(\begin{matrix}h,h,h-\tfrac12\\\tfrac12,c_h,2h-1\end{matrix};\tau u\right).
\end{equation}
Keeping $\normalB$ arbitrary, the remaining Laplace--Mellin transform in
\eqref{eq:B-second-integral} yields the Kamp\'e de F\'eriet representation
$F^{1:0:3}_{0:1:3}$ of \cite[Eq.~(3.27)]{Shimada}, with $v_{234}=\normalB^2$.
With $\normalB=0$ as well, all four points are collinear
and \eqref{eq:B-gamma} gives the block \eqref{eq:collinear}:
\begin{equation}\label{eq:B-four-collinear}
 \cF_h(\tau,0,0)=\hyp{4}{3}\!\left(
 \begin{matrix}h,h,h,h-\tfrac12\\\tfrac12,c_h,2h-1\end{matrix};\tau\right)
 =\tau^{-h}B_{+,h}(\tau).
\end{equation}
The special collinear blocks in Section~\ref{sec:free-limits} follow from
\cite[Eqs.~(67)--(68)]{BurchnallChaundy1941}.
With $x=\tau u/4$, Eq.~(67) gives at $h=1/2$
\begin{equation}\label{eq:B-half-diagonal}
 \Psi_2\!\left(\tfrac12;\tfrac12,\tfrac12;x,x\right)
 =\lim_{\varepsilon\to0}\hyp{1}{1}(\varepsilon;2\varepsilon;4x)
 =\tfrac12(1+e^{4x}).
\end{equation}
The joint limit is essential: $(\varepsilon)_n/(2\varepsilon)_n\to1/2$ for $n\ge1$.
At $h=1$, Eq.~(68) gives
\begin{equation}\label{eq:B-one-diagonal}
 \Psi_2\!\left(1;\tfrac12,\tfrac32;x,x\right)=\hyp{1}{1}(1;\tfrac32;4x).
\end{equation}
The remaining Laplace--Mellin transforms recover
\eqref{eq:free-collinear} and \eqref{eq:h-one}, respectively.
For the odd block at $h=1$, the sign insertion in \eqref{eq:B-odd-integral}
turns the first sign sum into a $\sinh$ term; Eq.~(67) then gives
\begin{equation}\label{eq:B-odd-one-diagonal}
 \Psi_2\!\left(\tfrac32;\tfrac32,\tfrac32;x,x\right)
 =\hyp{1}{1}(1;2;4x)=\frac{e^{4x}-1}{4x}.
\end{equation}
The corresponding Laplace--Mellin transform gives \eqref{eq:free-fermion}.

Returning to general $(\normalA,\normalB)$, the factor $4^p$ in \eqref{eq:B-mellin-convolution}
cancels the $4^{-p}$
from \eqref{eq:B-XZ} and the second argument of $\Psi_2$. Consequently
\begin{equation}\label{eq:B-mellin-triple}
 \cF_h=\sum_{p,\ell,k\ge0}\frac{\tau^p\normalA^{2\ell}\normalB^{2k}}{p!\ell!k!}
 \bigl[\Lambda_{p\ell k}^{(0)}+z\Lambda_{p\ell k}^{(1)}\bigr],
\end{equation}
with
\begin{equation}\label{eq:B-mellin-coefficients}
 \Lambda_{p\ell k}^{(j)}=(2h)^{2j}
 \frac{(h+j)_{p+\ell}(h+j)_{p+k}(2h-1+j)_{2p}}
 {(\tfrac12+j)_\ell(\tfrac12+j)_k(\tfrac12+j)_p(4h-2+j)_{2p}}.
\end{equation}
These are exactly \eqref{eq:triple}--\eqref{eq:lambda-coeff} 
with $q=h$ in the notation of Section~\ref{sec:solutions}.
At removable singularities the coefficients are understood by continuation of the
whole expression, as in Section~\ref{sec:even}.

\subsection{\texorpdfstring{Convergence and Kummer products from Gaussian moments}{Convergence and Kummer products from Gaussian moments}}\label{app:gaussian-moments}
To obtain the Kummer products directly, recombine the sign sums into an integral
over $\mathbb R^2$, using the signed variables
\begin{equation}\label{eq:B-scaling}
 \intx=a(x_2-x_1'),\qquad \inty=\widetilde a(x_4'-x_3).
\end{equation}
Here $\inty$ is $\rho_4$ times the positive integration variable used above
and ranges over $\mathbb R$.
Using \eqref{eq:B-positive-scaling}, the constants in \eqref{eq:B-split} and the Bessel series give
\begin{equation}\label{eq:B-normalization}
 W=\cP\,\tau^h\frac{\cI_h}{2^{2h-1}\pi^{3/2}\Gamma(2h-\tfrac12)},
 \qquad \cF_h=\frac{\cI_h}{\Gamma(h)^2},
\end{equation}
where
\begin{equation}\label{eq:B-integral}
 \cI_h=\int_{\mathbb R^2}d\intx\,d\inty\,|\intx\inty|^{2h-1}
 e^{-\intx^2-\inty^2-\sqrt\tau\,\intx\inty-2\normalA\intx+2\normalB\inty}\,
 \hyp{0}{1}\!\left(;c_h;\frac{\tau\intx^2\inty^2}4\right).
\end{equation}
For $h\ge1/2$, $0\le\tau<1$, and finite real $(\normalA,\normalB)$, absolute convergence
follows from $\hyp{0}{1}(;c_h;v^2/4)\le\cosh|v|$ and
\begin{equation}\label{eq:B-bound}
 \intx^2+\inty^2-2\sqrt\tau\,|\intx\inty|\ge(1-\sqrt\tau)(\intx^2+\inty^2).
\end{equation}
The same Gaussian bound dominates the integrand uniformly for
$(\normalA,\normalB)$ in compact subsets of $\mathbb C^2$. Thus the integral is entire in these variables, and its
spatial Taylor series converges locally uniformly.
These bounds justify the termwise integrations in both evaluations; other parameters
are then treated by continuation.
For $h\ge1/2$, $0\le\tau<1$, and real $(\normalA,\normalB)$, every factor of the
integrand in \eqref{eq:B-integral} is nonnegative, so $U_{+,h}=\tau^h\cF_h\ge0$.
The mixed kernel in \eqref{eq:B-integral} is
$\cK_h(\sqrt\tau\,\intx\inty)$, where
\begin{equation}\label{eq:B-mixed-kernel}
 \cK_h(\omega)=e^{-\omega}\hyp{0}{1}(;c_h;\omega^2/4)
 =\sum_{n\ge0}\sum_{j=0}^1(-1)^jS_n^{(j)}\omega^{2n+j}.
\end{equation}
The coefficients are given by the convolution already evaluated in \eqref{eq:B-mellin-convolution}:
\begin{equation}\label{eq:B-finite-sum}
 S_n^{(j)}=4^{-n}\sum_{M+k=n}\frac1{M!k!(\tfrac12+j)_M(c_h)_k}
 =\frac{(c_h+n+j-\tfrac12)_n}{(c_h)_n(2n+j)!}.
\end{equation}
Here $(2M+j)!=4^M M!(\tfrac12+j)_M$ converts the exponential coefficient into this form.
The required one-dimensional moments follow from the Gamma integral \eqref{eq:B-gamma}:
\begin{equation}\label{eq:B-moments}
 \begin{aligned}
 \int_{\mathbb R}|x|^{2h-1}x^{2n}e^{-x^2+2dx}dx
 &=\Gamma(h+n)\hyp{1}{1}\!\left(h+n;\tfrac12;d^2\right),\\
 \int_{\mathbb R}|x|^{2h-1}x^{2n+1}e^{-x^2+2dx}dx
 &=2d\,\Gamma(h+n+1)\hyp{1}{1}\!\left(h+n+1;\tfrac32;d^2\right).
 \end{aligned}
\end{equation}
The first follows by expanding $\cosh(2dx)$ after $u=x^2$; half its $d$-derivative gives
the second. At $d=-\normalA$ and $d=\normalB$, the minus sign from the odd $\intx$ moment
cancels $(-1)^j$ from the kernel, giving
\begin{equation}\label{eq:B-coefficients}
 b_n^{(+)}(h)=(h)_n^2S_n^{(0)},\qquad y_n^{(+)}(h)=4(h)_{n+1}^2S_n^{(1)}.
\end{equation}
Using \eqref{eq:pochhammer} reduces these to \eqref{eq:coefficients}.
The two moment functions are $E_{h+n}$ and $O_{h+n+1/2}$ up to their Gamma factors,
so this evaluation gives the Kummer-product sum \eqref{eq:full-block} directly.

For the odd primary, let $d\mu_h$ denote the full integrand measure defining $\cI_h$ in
\eqref{eq:B-integral}. The antisymmetric relative kernel inserts
$-\operatorname{sgn}(\intx\inty)$, with the sign fixed by
$\intx\propto x_2-x_1'$ and $\inty\propto x_4'-x_3$ and the ordering in
\eqref{eq:correlator}. The normalized odd block is
\begin{equation}\label{eq:B-odd-integral}
 U_{-,h}=\frac{\tau^h}{\Gamma(h+\tfrac12)^2}
 \int_{\mathbb R^2}[-\operatorname{sgn}(\intx\inty)]\,d\mu_h.
\end{equation}
Since the inserted sign has modulus one, the bound \eqref{eq:B-bound} also applies.
Expand the integrand of \eqref{eq:B-odd-integral} in powers of $\sqrt\tau$ at fixed
$(\normalA,\normalB)$. The constant term in this expansion factorizes into signed one-dimensional
moments and gives $4\tau^hO_h(\normalA)O_h(\normalB)=4\tau^h\normalA\normalB+\cdots$,
from which we obtain the leading term $Y=4\tau^{h-1/2}$ of the rescaled mixed seed.
This contribution vanishes on the collinear locus, where the first nonzero term of
the mixed kernel is linear in $\sqrt\tau$ and gives the leading term $B=\tau^{h+1/2}$.
This verifies the primary parity and the normalization of
\eqref{eq:odd-seeds}. Applying the same finite sums \eqref{eq:B-finite-sum} gives
\begin{equation}\label{eq:B-odd-coefficients}
 b_n^{(-)}(h)=(h+\tfrac12)_n^2 S_n^{(1)},\qquad
 y_n^{(-)}(h)=(h+\tfrac12)_n^2 S_n^{(0)},
\end{equation}
which are precisely \eqref{eq:odd-coefficients} 
at branch parameter $q=h$ in the notation of Section~\ref{sec:solutions}.
 
The common prefactor in
\eqref{eq:B-normalization}, together with the different gamma denominator in
\eqref{eq:B-odd-integral}, gives \eqref{eq:odd-normalization}.

For the even block at $h=1/2$, the mixed kernel \eqref{eq:B-mixed-kernel} simplifies to
\begin{equation}\label{eq:B-free}
 \cK_{1/2}(\omega)=e^{-\omega}\cosh\omega=\tfrac12(1+e^{-2\omega}).
\end{equation}
Thus \eqref{eq:B-integral} is the sum of an uncoupled Gaussian and a Gaussian with
quadratic matrix $\left(\begin{smallmatrix}1&\sqrt\tau\\\sqrt\tau&1\end{smallmatrix}\right)$ and source
$(-\normalA,\normalB)$. Evaluating them gives \eqref{eq:free-full}.

\subsection{\texorpdfstring{The hard-core boson block and Brownian-bridge sign correlations}{The hard-core boson block and Brownian-bridge sign correlations}}\label{app:unlike}
We first identify the two-particle kernel used in Section~\ref{sec:dla}.
Let an $A$ particle start at $y_1$ and a $B$ particle at $y_2$ at time $t_2$.
With the heat kernel $k_\ell$ defined in \eqref{eq:bridge-density}, consider a determinant
\begin{equation}\label{eq:B-km-determinant}
 \begin{aligned}
 \mathcal D(z_1,z_2;y_1,y_2)
 &=\det\!\bigl[k_{t_{32}}(z_i-y_j)\bigr]_{i,j=1,2}\\
 &=k_{t_{32}}(z_1-y_1)k_{t_{32}}(z_2-y_2)
   -k_{t_{32}}(z_1-y_2)k_{t_{32}}(z_2-y_1).
 \end{aligned}
\end{equation}
By the Karlin--McGregor formula \cite{KarlinMcGregor}, the joint density for
the particles to reach $z_1,z_2$ at $t_3$ without meeting is $\mathcal D$ when
$(y_1-y_2)(z_1-z_2)>0$, and zero otherwise. 
The two assignments of the final
positions to the species give the symmetrized kernel
\begin{equation}\label{eq:B-hard-core-kernel}
 K_{\rm hc}^{(2)}(z_1,z_2;y_1,y_2;t_{32})
 =\operatorname{sgn}(y_1-y_2)\operatorname{sgn}(z_1-z_2)
   \mathcal D(z_1,z_2;y_1,y_2).
\end{equation}
For either initial ordering, the unique assignment that preserves the particle order gives the nonnegative density on the corresponding ordering sector.

Equation~\eqref{eq:B-hard-core-kernel} is the two-particle hard-core boson kernel
obtained from Girardeau's Bose--Fermi mapping \cite[Eqs.~(3)--(4)]{Girardeau}.

Alternatively, using $I_{1/2}(w)=(e^w-e^{-w})/\sqrt{2\pi w}$, we obtain
${\left.K_2\right|_{\nu=1/2}=|\mathcal D|=K_{\rm hc}^{(2)}}$
from the Calogero kernel in \eqref{eq:B-kernels}.
Convolving it with the one-particle kernels in the outer intervals gives the
unlike-species contribution in \eqref{eq:dla}.

For collinear insertions $x_i=0$, the direct term in the convolution of
\eqref{eq:B-hard-core-kernel} with the outer heat kernels factorizes into two
centered Gaussian sign integrals and vanishes. The crossed term is
$k_{t_{32}}(0)k_{t_{41}}(0)$ times the sign correlation of a centered Brownian bridge
at $t_2$ and $t_3$. 
As the two bridge values have correlation coefficient $\sqrt\tau$,
Sheppard's formula for Gaussian quadrant probabilities
\cite[Sec.~27]{Sheppard}  
gives
\footnote{
 In Sheppard's notation, each of the two same-sign quadrants contains $P$ observations, and each of the two opposite-sign quadrants contains $R$, with $D=\pi R/(P+R)$.
Writing the correlation coefficient as $\rho=\cos D$, we therefore obtain
$\mathbb E[\operatorname{sgn}X\operatorname{sgn}Y]=(2P-2R)/(2P+2R)=1-2D/\pi=(2/\pi)\arcsin\rho$
for centered $X,Y$. 
} 
their sign correlation as
$(2/\pi)\arcsin\sqrt\tau$.
For this correlation, see \cite[Eq.~(5.34)]{Godreche} for the explicit
Brownian-bridge formula and \cite[Eqs.~(48) and (55)]{BCS}
for a direct Gaussian-integral evaluation.
After dividing by the kinematic prefactor and normalizing the
leading collinear coefficient to one, we obtain
$B(\tau)=\sqrt{\tau/(1-\tau)}\arcsin\sqrt\tau=B_{+,1}(\tau)$.
This is the collinear input used in Section~\ref{sec:dla-signed}.

At $h=1$, the mixed kernel simplifies to
$\cK_1(\omega)=e^{-\omega}\sinh\omega/\omega=(1-e^{-2\omega})/(2\omega)$,
with its continuous value $1$ at $\omega=0$. The corresponding factor in the
integrand of \eqref{eq:B-integral} is
\begin{equation}\label{eq:B-unlike-kernel-split}
 |\intx\inty|\,\cK_1(\sqrt\tau\,\intx\inty)
 =\frac{\operatorname{sgn}(\intx\inty)}{2\sqrt\tau}
 \left(1-e^{-2\sqrt\tau\,\intx\inty}\right).
\end{equation}
The integral splits into an uncoupled and a coupled Gaussian integral, both weighted by
$\operatorname{sgn}(\intx\inty)$. They give the direct and the crossed term,
respectively. Using the Lorentz-boosted normal coordinates
(see Appendix~\ref{app:walls}),
\begin{equation}\label{eq:B-alpha-beta}
 \alpha=\frac{\normalA+\sqrt\tau\,\normalB}{\sqrt{1-\tau}},\qquad
 \beta=\frac{\sqrt\tau\,\normalA+\normalB}{\sqrt{1-\tau}},
\end{equation}
we obtain the full hard-core boson block
\begin{equation}\label{eq:B-unlike-closed}
 U_{+,1}=\frac{\pi\sqrt\tau}{2}\Bigl[\frac{e^{Q_\tau(\normalA,\normalB)}}{\sqrt{1-\tau}}\,
 G_\tau(\alpha,\beta)-e^{\normalA^2+\normalB^2}\operatorname{erf}(\normalA)
 \operatorname{erf}(\normalB)\Bigr].
\end{equation}
Here $Q_\tau$ is the quadratic form \eqref{eq:gaussian-form}, and
$G_\tau(\alpha,\beta)=\mathbb E[\operatorname{sgn}(X-\alpha)\operatorname{sgn}(Y-\beta)]$
for centered jointly Gaussian variables $X,Y$ with variance $1/2$ and correlation $\sqrt\tau$.
An alternative interpretation is obtained by considering a Brownian bridge $X_{\rm br}$
from $(t_1,x_1)$ to $(t_4,x_4)$.
Its values at $t_2,t_3$ have the same correlation $\sqrt\tau$.
The coordinates $\alpha,\beta$ in \eqref{eq:B-alpha-beta} are the
negatives of the normalized deviations of the insertion positions
from the mean bridge path as in \eqref{eq:signed-cross-terms}.
The expectation of
$\operatorname{sgn}(X_{\rm br}(t_2)-x_2)
 \operatorname{sgn}(X_{\rm br}(t_3)-x_3)$
is $G_\tau(-\alpha,-\beta)=G_\tau(\alpha,\beta)$,
where the equality follows from the symmetry of the centered Gaussian
distribution under simultaneous sign reversal.
Keeping $\alpha,\beta$ fixed and integrating the Gaussian differential identity
\cite{Plackett} (see also \cite[Eqs.~(2.5)--(2.8)]{Meyer})
over the correlation coefficient gives
\begin{equation}\label{eq:B-sign-integral}
 G_\tau(\alpha,\beta)
 =\operatorname{erf}(\alpha)\operatorname{erf}(\beta)
 +\frac2\pi\int_0^{\sqrt\tau}
 \frac{\exp\!\left[-\frac{\alpha^2+\beta^2-2r\alpha\beta}{1-r^2}\right]}
 {\sqrt{1-r^2}}\,dr.
\end{equation}
At $\alpha=\beta=0$, we recover $G_\tau(0,0)=(2/\pi)\arcsin\sqrt\tau$
and hence \eqref{eq:h-one}.

The identity $\operatorname{erf}(x)=(2x/\sqrt\pi)e^{-x^2}\hyp{1}{1}(1;\tfrac32;x^2)$
and \eqref{eq:invisible-lift} identify the direct term of \eqref{eq:B-unlike-closed}
with $-2U_{\mathrm{inv}}$.
Denoting the crossed term by $C_\tau$, we obtain
\begin{equation}\label{eq:sign-structure}
 U_{+,1}=C_\tau-2U_{\mathrm{inv}},\qquad U_{-,1/2}=C_\tau+2U_{\mathrm{inv}}.
\end{equation}
The second relation follows from \eqref{eq:B-odd-integral} for the FTG block\footnote{
    For two particles, the FTG propagator is
$K_{\rm FTG}^{(2)}(\mathbf z;\mathbf y;t)=
\operatorname{sgn}(z_1-z_2)\operatorname{sgn}(y_1-y_2)
\operatorname{perm}[k_t(z_i-y_j)]$,
the Bose--Fermi dual of the free-boson kernel, where
$\operatorname{perm}M=M_{11}M_{22}+M_{12}M_{21}$.
Here the kernel is defined using the initial state
$\overline\Psi(y_1)\overline\Psi(y_2)|0\rangle$.
With $(y_1,y_2)=(x_1',x_2)$, the ordering in \eqref{eq:correlator}
instead gives
$\overline\Psi(y_2)\overline\Psi(y_1)
=-\overline\Psi(y_1)\overline\Psi(y_2)$,
so the convolution contains $-K_{\rm FTG}^{(2)}$,
in agreement with \eqref{eq:B-odd-integral}.}
at $h=1/2$: compared with \eqref{eq:B-unlike-kernel-split}, 
the factor $-\operatorname{sgn}(\intx\inty)\,\cK_{1/2}$ from \eqref{eq:B-free} 
keeps the coupled term and reverses the uncoupled one. 

\subsection{\texorpdfstring{Hille--Hardy decomposition for the state--operator map}{Hille--Hardy decomposition for the state--operator map}}\label{app:hille-hardy}
We use the separations $r,s>0$ and the center-of-mass displacement $R$
of Section~\ref{sec:calogero-model}, 
with the untrapped time interval ${\mathcal T}=t_{32}$
in the pairwise equal-time configuration.
The Hamiltonian for the relative motion in \eqref{eq:equal-time-evolution} is
$H_{\nu,0}=-\partial_r^2+(\nu^2-\tfrac14)/r^2$.
For $\nu\ge-1/2$ we choose the scale-invariant contact condition
selecting the pure $r^{\nu+1/2}$ branch,
excluding the $r^{1/2-\nu}$ branch for $\nu\ne0$ and the
$r^{1/2}\log r$ branch at $\nu=0$. Its propagator is
\begin{equation}\label{eq:B-relative-kernel}
 K_{\nu,0}(r,s;{\mathcal T})=\frac{\sqrt{rs}}{2{\mathcal T}}
 \exp\!\left[-\frac{r^2+s^2}{4{\mathcal T}}\right]
 I_\nu\!\left(\frac{rs}{2{\mathcal T}}\right).
\end{equation}
Multiplication by the center-of-mass kernel
$e^{-R^2/{\mathcal T}}/\sqrt{\pi {\mathcal T}}$ gives the two-particle kernel in \eqref{eq:B-kernels}.
To relate this propagator to the state--operator map \cite{NishidaSon},
we introduce a harmonic trap of frequency $\omega>0$,
$H_{\nu,\omega}=H_{\nu,0}+\omega^2r^2/4$.
Its normalized Laguerre eigenfunctions and energies are
\begin{equation}\label{eq:laguerre-modes}
 \begin{aligned}
 \varphi_n(r)&=\left[\frac{\omega^{\nu+1}n!}{2^\nu\Gamma(n+\nu+1)}\right]^{1/2}
 r^{\nu+1/2}e^{-\omega r^2/4}L_n^{(\nu)}(\omega r^2/2),\\
 E_n&=\omega(2n+\nu+1),\qquad n=0,1,\ldots.
 \end{aligned}
\end{equation}
As discussed in Section~\ref{sec:calogero-model}, including the
center-of-mass ground-state energy gives
$(E_n+\omega/2)/\omega=\Delta+2n$, where $\Delta=\nu+3/2$.
This identifies the relative-motion descendant tower whose contribution
to the propagator is summed below; the center-of-mass contribution
is treated separately.

Let $u>0$ denote the Euclidean evolution time of the trapped system.
The Hille--Hardy formula \cite[Eq.~(18.18.27)]{DLMF}, with Laguerre index
$\nu$ and summation parameter $e^{-2\omega u}$, gives its propagator:
\begin{equation}\label{eq:hille-hardy-kernel}
 \begin{aligned}
 K_{\nu,\omega}(r,s;u)
 &=\sum_{n=0}^\infty e^{-E_nu}\varphi_n(r)\varphi_n(s)\\
 &=\frac{\omega\sqrt{rs}}{2\sinh(\omega u)}
 \exp\!\left[-\frac{\omega(r^2+s^2)}4\coth(\omega u)\right]
 I_\nu\!\left(\frac{\omega rs}{2\sinh(\omega u)}\right).
 \end{aligned}
\end{equation}
At fixed $u$, the $\omega\to0$ limit of \eqref{eq:hille-hardy-kernel}
recovers \eqref{eq:B-relative-kernel} with ${\mathcal T}=u$.
To relate this resummed trapped propagator to the untrapped one at finite $\omega$, we use
\begin{equation}\label{eq:trap-kernel-map}
 K_{\nu,\omega}(r,s;u)
 =\frac{e^{-\omega r^2\tanh(\omega u)/4}}{\sqrt{\cosh(\omega u)}}
 K_{\nu,0}\!\left(\frac{r}{\cosh(\omega u)},s;
 \frac{\tanh(\omega u)}{\omega}\right).
\end{equation}
This identity follows directly from the two Bessel expressions.
It relates the trapped and untrapped evolution through
${\mathcal T}=t_{32}=\tanh(\omega u)/\omega$, together with the rescaling
$r\mapsto r/\cosh(\omega u)$.
For a related construction of the many-body Calogero propagator
from untrapped eigenfunctions, see \cite{MelinLangmann}.

At $\nu=\mp1/2$, the Laguerre modes correspond to the even and odd Hermite
sectors of Mehler's formula \cite[Eqs.~(18.7.19)--(18.7.20)]{DLMF}.
The same Mehler structure appears in the expansion after \eqref{eq:free-full},
but with a different physical role: here it describes propagation in the
relative coordinate, whereas there it organizes the dependence of the
four-point function on analytically continued shape variables.

\section{\texorpdfstring{Four-point invariants: walls, channels, and square-root branches}{Four-point invariants: walls, channels, and square-root branches}}\label{app:invariants}
This appendix examines three properties of the signed coordinates introduced in
Section~\ref{sec:kinematics}, namely, how they determine the remaining invariants, how the insertions can be
relabeled, and how the coordinates behave when a square-root branch changes.
The first property gives the walls of Fig.~\ref{fig:residuals}(b), and the second
relates the charged channel to the two neutral channels.
The third shows that, on each generic solution branch, the two Taylor sectors
\eqref{eq:parity-decomposition} acquire opposite multipliers under continuation around $\tau=0$
at fixed $(\normalA,\normalB)$.

\subsection{\texorpdfstring{Determinant identities, a Lorentz boost, and the wall cross-ratio}{Determinant identities, a Lorentz boost, and the wall cross-ratio}}\label{app:walls}
The four rows $(1,t_i,x_i)$ form a $4\times3$ matrix whose maximal minors are the
$3\times3$ determinants $A_{ijk}$ defined in \eqref{eq:z}.
Geometrically, $A_{ijk}$ is twice the signed area of the triangle with vertices
$(t_i,x_i)$, $(t_j,x_j)$, and $(t_k,x_k)$ \cite{Shimada}. The signed normal coordinates
retain this orientation information after normalization by the time separations.
We denote the minor obtained by
omitting row $i$ by $A_{\hat\imath}$.
Cofactor expansion gives the vector identity
$\sum_{i=1}^4(-1)^{i+1}(1,t_i,x_i)\,A_{\hat\imath}=(0,0,0)$.
The first two
components give $A_{234}-A_{134}+A_{124}-A_{123}=0$ and
$t_1A_{234}-t_2A_{134}+t_3A_{124}-t_4A_{123}=0$. Solving for $A_{124}$ and $A_{134}$ gives
\begin{equation}\label{eq:determinants}
 A_{124}=\frac{t_{42}A_{123}+t_{21}A_{234}}{t_{32}},\qquad
 A_{134}=\frac{t_{43}A_{123}+t_{31}A_{234}}{t_{32}}.
\end{equation}
Dividing by the normalizations of \eqref{eq:ab} and using \eqref{eq:tau} gives the
signed version of \eqref{eq:cross-terms},
\begin{equation}\label{eq:signed-cross-terms}
 \begin{aligned}
 \alpha=\frac{\normalA+\sqrt\tau\,\normalB}{\sqrt{1-\tau}}&=\frac{A_{124}}{\sqrt{2t_{21}t_{42}t_{41}}}
 =-\frac{\rho_{2|14}}{\sqrt{2t_{21}t_{42}/t_{41}}},\\
 \beta=\frac{\sqrt\tau\,\normalA+\normalB}{\sqrt{1-\tau}}&=\frac{A_{134}}{\sqrt{2t_{31}t_{43}t_{41}}}
 =-\frac{\rho_{3|14}}{\sqrt{2t_{31}t_{43}/t_{41}}},
 \end{aligned}
\end{equation}
where $\rho_{2|14}$ and $\rho_{3|14}$ are the residuals of insertions 2 and 3 from the
chord $14$, defined as in \eqref{eq:rho}. Squaring these relations and using $v_{ijk}=A_{ijk}^2/(2t_{ji}t_{kj}t_{ki})$
gives \eqref{eq:cross-terms}.
Here $\alpha$ and $\beta$ are the variables defined in \eqref{eq:B-alpha-beta}. 
In particular, $\alpha^2=v_{124}$ and $\beta^2=v_{134}$.
Subtracting the two relations in \eqref{eq:cross-terms} and using \eqref{eq:squares},
one finds $v_{124}-v_{134}-(v_{123}-v_{234})=0$.
This identity has a simple explanation: with
$E_{ij}=x_{ij}^2/(2t_{ij})$, antisymmetric in $ij$, each invariant is
$v_{ijk}=E_{ji}+E_{kj}+E_{ik}$, as in \eqref{eq:slope}. In the alternating sum over the four
faces of the tetrahedron with vertices at the insertions, each edge occurs twice with opposite
orientations.
In signed coordinates, this identity reads
$\alpha^2-\beta^2=\normalA^2-\normalB^2$.
Correspondingly, the transformation in \eqref{eq:signed-cross-terms} has the form
of a Lorentz boost! With
$\tanh\varpi=\sqrt\tau$, it becomes diagonal in the sum and difference coordinates:
$\alpha+\beta=e^{\varpi}(\normalA+\normalB)$ and
$\alpha-\beta=e^{-\varpi}(\normalA-\normalB)$.
Writing \eqref{eq:signed-cross-terms} as $(\alpha,\beta)^{\mathsf T}=L(\normalA,\normalB)^{\mathsf T}$,
the covariance matrix \eqref{eq:covariance} becomes
$L\Sigma_\tau L^{\mathsf T}=\tfrac12\bigl(\begin{smallmatrix}1&\sqrt\tau\\\sqrt\tau&1\end{smallmatrix}\bigr)$.
The boost therefore turns the correlation $-\sqrt\tau$ of $(\normalA,\normalB)$ into the
correlation $+\sqrt\tau$ of the residuals from the chord $14$, that is, of the bridge positions
in the footnote of Section~\ref{sec:residuals}.

The four walls are the zero sets of $A_{123}$, $A_{234}$, $A_{124}$, and $A_{134}$,
each corresponding to three collinear insertions in the $(t,x)$-plane.
In particular, $\alpha=0$ and $\beta=0$ are the walls $A_{124}=0$ and $A_{134}=0$. In the
normal plane the four walls are the lines through the origin with covectors
$\ell_{123}=(1,0)$, $\ell_{234}=(0,1)$, $\ell_{124}=(1,\sqrt\tau)$, and
$\ell_{134}=(\sqrt\tau,1)$, whose projective cross-ratio is
\begin{equation}\label{eq:wall-cross-ratio}
 \frac{\det(\ell_{123},\ell_{124})\det(\ell_{234},\ell_{134})}
 {\det(\ell_{123},\ell_{134})\det(\ell_{234},\ell_{124})}=\tau.
\end{equation}
The time cross-ratio $\tau$ therefore also measures the projective
arrangement of the walls.

\subsection{\texorpdfstring{Relabelings, channels, and the cross-ratio}{Relabelings, channels, and the cross-ratio}}\label{app:relabellings}
The four insertions can be split into two pairs in three ways, $12|34$, $14|23$, and
$13|24$. In the convention of \cite{Shimada}, $12|34$ is the charged $S$ channel,
while $13|24$ and $14|23$ are the neutral $T$ and $U$ channels.
The corresponding products of time gaps obey
\begin{equation}\label{eq:ptolemy}
 t_{31}t_{42}=t_{21}t_{43}+t_{32}t_{41},
\end{equation}
so that $\tau$ and $1-\tau$ in \eqref{eq:tau} are obtained by dividing the first and second
terms on the right-hand side by $t_{31}t_{42}$, respectively. The charged $S$ channel and the neutral $U$ and $T$ channels
correspond to $\tau=0$, $1$, and $\infty$, respectively,
where the corresponding product of time gaps vanishes. The Klein four-group
$V_4=\{e,(12)(34),(13)(24),(14)(23)\}$ of relabelings of the insertions
consists of the relabelings that map each of the three pairings to itself. It fixes each of the three products, hence fixes $\tau$ and
the wall cross-ratio \eqref{eq:wall-cross-ratio}, and it permutes the four walls by the same
double transpositions of the omitted labels. The quotient $S_4/V_4\simeq S_3$ permutes the
pairings and produces the six values $\tau$, $1-\tau$, $1/\tau$, $1/(1-\tau)$, $\tau/(\tau-1)$,
and $(\tau-1)/\tau$ (see, e.g., \cite[Lemma~4.4]{BambergPenttila}).

The exchanges $(12)$ and $(34)$ preserve the assignment of $\overline\Psi$ and
$\Psi$ and the charged channel $12|34$, but take the ordered chart to other time-ordering regions.
The time-ordered correlator obeys the corresponding exchange relations, with a minus
sign for each exchange of fermionic fields.

To express these exchange relations in terms of the reduced blocks, one must
also transform the prefactor and the signed coordinates, keeping track of their
square-root branches. For $(12)(34)$, taking square roots of the negative
normalization factors with positive imaginary part gives
$(\normalA',\normalB')=i(\alpha,\beta)$. This coordinate transformation is $i$ times
the Lorentz boost in \eqref{eq:signed-cross-terms}.
The factor $i$ reflects the antisymmetry of $v_{ijk}$:
$v_{214}=-v_{124}$ and $v_{143}=-v_{134}$.
With these square roots, \eqref{eq:z} continues to hold after the relabeling
only if the sign of $\sqrt\tau$ is also reversed, although $\tau$ itself is unchanged.
By the sign changes of Appendix~\ref{app:branches}, this is equivalent to keeping
$\sqrt\tau$ and reversing $\normalB'$.
More general relabelings relate the three channels and the
regions near $\tau=0$, $1$, and $\infty$.
We leave the transformation of the blocks under these relabelings and the
computation of the resulting connection matrices to a future crossing analysis.

\subsection{\texorpdfstring{Square-root branches and the monodromy of the two sectors}{Square-root branches and the monodromy of the two sectors}}\label{app:branches}
The signed description $(\tau;\normalA,\normalB)$ involves three square roots: the
time-gap roots in $\normalA$ and $\normalB$, and $\sqrt\tau$. The squares $\normalA^2$,
$\normalB^2$, and the product $z=\sqrt\tau\,\normalA\normalB$ are rational, while the three factors
separately are not. The sign changes
$(\sqrt\tau,\normalA,\normalB)\mapsto(\epsilon_0\sqrt\tau,\epsilon_1\normalA,\epsilon_2\normalB)$
with $\epsilon_0\epsilon_1\epsilon_2=1$ form a group of order four. It leaves $z$, $v_{123}$,
$v_{234}$, and \eqref{eq:cross-terms} unchanged, so it acts trivially on all rational invariants.
Its elements are the identity, spatial reflection, and two square-root branch
changes that reverse $\sqrt\tau$. The group $V_4$ of insertion relabelings in
Appendix~\ref{app:relabellings} preserves $\tau$ but acts nontrivially on the spatial
invariants.

On analytic continuation away from the ordered chart, the second root
$\sqrt{1-\tau}$, which enters the frame \eqref{eq:frame} and the free
block \eqref{eq:free-full}, can also change sign. The half-unit shift between the two primary parities at $\tau=0$
reflects the Schr\"odinger scaling weights of spatial and temporal descendants, as in
\eqref{eq:first-states}. Together with the collision exponent $-\tfrac12$ at $\tau=1$ in
\eqref{eq:collision}, this makes it natural to keep track of both $\sqrt\tau$ and
$\sqrt{1-\tau}$. The pair
$(c,s)=(\sqrt{1-\tau},\sqrt\tau)$ satisfies $c^2+s^2=1$. Over the $\tau$-plane with $0$, $1$, and
$\infty$ removed it has four sheets; a loop around $\tau=0$ reverses $s$, and a loop around
$\tau=1$ reverses $c$. The temporal part of the ordered chart corresponds to the open arc
$c,s>0$ of the real unit circle.
With the radial coordinate $\rho$ of Section~\ref{sec:two-body-extensions},
$\tau=4\rho/(1+\rho)^2$, one has $c=(1-\rho)/(1+\rho)$ and $s=2\sqrt\rho/(1+\rho)$. Both roots
are rational in $\sqrt\rho$, and $0<\sqrt\rho<1$ on the ordered chart. The loops around
$\tau=0$ and $\tau=1$ act as $\sqrt\rho\mapsto-\sqrt\rho$ and $\sqrt\rho\mapsto1/\sqrt\rho$,
respectively.\footnote{The six points $\sqrt\rho=0,\infty,\pm1,\pm i$ lying above
$\tau=0,1,\infty$ form the vertices of a regular octahedron on the Riemann sphere. 
The M\"obius transformations preserving this set form the octahedral group $S_4$
(see, e.g., \cite[Part~I, Chaps.~I--II]{Klein}).
Through $\tau=4\rho/(1+\rho)^2$, they induce the anharmonic $S_3$ action on $\tau$.
The transformations that leave $\tau$ unchanged form the normal Klein subgroup
generated by $\sqrt{\rho}\mapsto-\sqrt{\rho}$ and $\sqrt{\rho}\mapsto1/\sqrt{\rho}$.
On $(c,s)$, these act as $(c,s)\mapsto(c,-s)$ and $(c,s)\mapsto(-c,s)$, respectively.
Thus this subgroup consists of the independent sign changes of the two square roots,
and $S_4/V_4\simeq S_3$.}
Although $c$ and $s$ are single-valued on this cover, the factor
$\tau^p=[4\rho/(1+\rho)^2]^p$ remains multivalued for generic $p$.

The solutions of Section~\ref{sec:solutions} are sensitive to the first of these loops in
a simple way. On each of the four generic local branches of Section~\ref{sec:shadow}, the lift
\eqref{eq:lift} shows that $F$ and $G$ in \eqref{eq:parity-decomposition} are $\tau^p$ times power
series in $\tau$, with coefficients analytic in $(v_{123},v_{234})$ and a common exponent $p$: $p=q$ for the
even primary, by \eqref{eq:full-block}, and $p=q-\tfrac12$ for the odd primary, by
\eqref{eq:odd-full}. At fixed $(\normalA,\normalB)$ these power series converge absolutely and locally uniformly for
$|\tau|<1$: the coefficients \eqref{eq:coefficients} and \eqref{eq:odd-coefficients} grow at most
polynomially in $n$, while for large $\lambda$ the Kummer functions \eqref{eq:even-odd} grow only as
$\exp\bigl(O(\sqrt\lambda)\bigr)$ \cite[Eq.~(13.8.12)]{DLMF}. The power series multiplying $\tau^p$ are even in $\sqrt\tau$,
whereas the factor $z$ is odd. Continuing $\tau$ once positively around $0$ at fixed $(\normalA,\normalB)$ inside this disk
therefore gives
\begin{equation}\label{eq:sector-monodromy}
 U(\tau;\normalA,\normalB)=F+zG\;\longmapsto\;e^{2\pi ip}\,(F-zG)=e^{2\pi ip}\,U(\tau;\normalA,-\normalB).
\end{equation}
The two sectors therefore acquire the multipliers $e^{2\pi ip}$ and $-e^{2\pi ip}$. The
continuation also reverses $\sqrt\tau$ in \eqref{eq:reduced}, so it leads to the sheet with
$(c,s)$ replaced by $(c,-s)$.
Lemma~\ref{lem:lift} is formulated on a fixed branch of $\sqrt\tau$, while
\eqref{eq:sector-monodromy} describes the continuation of the generic block solutions to that sheet.
The monodromy formula \eqref{eq:sector-monodromy} follows from the stated power-series form of the solutions.
At resonant parameters, a complete local basis may also require logarithmic solutions,
as at $h=3/4$ (Section~\ref{sec:shadow}).
The monodromy formula remains valid for the nonlogarithmic solutions,
while logarithmic solutions acquire additional contributions from other basis solutions under continuation.

\clearpage

\end{document}